%% file: lics25.tex
\documentclass[conference]{IEEEtran}
\IEEEoverridecommandlockouts

\usepackage{amsmath}
\usepackage{amsthm}

\usepackage{mathtools}
\usepackage{enumitem}
\usepackage{cite}
\usepackage{amssymb,amsfonts}
\usepackage{algorithmic}
\usepackage{graphicx}
\usepackage{textcomp}
\usepackage{xcolor}

\usepackage{physics2}
\usephysicsmodule{ab}
\usepackage{ stmaryrd }
\usepackage{todonotes}
\usepackage{bussproofs}
\usepackage{tikz-cd}
\usepackage{url}
\usetikzlibrary {arrows.meta,bending,positioning,graphs} 
\input{common.tex}
\def\BibTeX{{\rm B\kern-.05em{\sc i\kern-.025em b}\kern-.08em
    T\kern-.1667em\lower.7ex\hbox{E}\kern-.125emX}}

\newenvironment{bprooftree}
  {\leavevmode\hbox\bgroup}
  {\DisplayProof\egroup}

\newcommand{\Ty}{\mathrm{Ty}}
\newcommand{\Cl}{\mathrm{Cl}}
\newcommand{\ev}{\mathrm{ev}}

\newcommand{\source}{\mathsf{src}}
\newcommand{\target}{\mathsf{tgt}}

\newcommand{\gen}[1]{\underline{#1}}
\newcommand{\multiplicity}[2]{\mathfrak{m}_{#1}\ab(#2)}
\newcommand{\pr}{\mathsf{pr}}

\newcommand{\vareqn}{e}

\newcommand{\pp}{{\mathrm{pp}}}
\newcommand{\fp}{\mathfrak{p}}

\begin{document}

\title{Homological Invariants of\\ Higher-Order Equational Theories}
\author{\IEEEauthorblockN{Mirai Ikebuchi}
\IEEEauthorblockA{\textit{Graduate School of Informatics} \\
\textit{Kyoto University}\\
Kyoto, Japan\\
ikebuchi@fos.kuis.kyoto-u.ac.jp}
%\and
%\and
%\and
%\and
%\and
}

\maketitle

\begin{abstract}
Many first-order equational theories, such as the theory of groups or boolean algebras, can be presented by a smaller set of axioms than the original one.
Recent studies showed that a homological approach to equational theories gives us inequalities to obtain lower bounds on the number of axioms.

In this paper, we extend this result to higher-order equational theories.
More precisely, we consider simply typed lambda calculus with product and unit types and study sets of equations between lambda terms.
Then, we define homology groups of the given equational theory and show that a lower bound on the number of equations can be computed from the homology groups.
\end{abstract}


\section{Introduction}

It is known that the set of group axioms
\[
(x\cdot y) \cdot z = x \cdot (y \cdot z),\quad
x \cdot e = x,\quad
x \cdot x^{-1} = e
\]
is equivalent to the following set consisting of two equational axioms:
\begin{align*}
w \cdot (((x^{-1} \cdot (w^{-1}\cdot y))^{-1} \cdot z)\cdot (x\cdot z)^{-1})^{-1} = y,\,
x\cdot x^{-1} = e.
\end{align*}
Moreover, it is proved that there is no single equational axiom in function symbols $\cdot$, ${}^{-1}$, $e$ that is equivalent to the group axioms \cite{tarski1968,neumann1981,kunen1992}.
That is, the set with the two equations above is minimum in the sense of the number of equations in $\cdot$, ${}^{-1}$, $e$.

In \cite{mm16,ikebuchi22}, it is shown that homological and homotopical algebraic methods provide lower bounds on the number of equations for given sets of (first-order) equational axioms, not just group axioms.
The lower bound is algorithmically computable if the set of equations is a finite and complete (i.e., confluent and terminating) term rewriting system.
In particular, the nonexistence of a single group axiom immediately follows from a more general theorem given in \cite{ikebuchi22}.

In this paper, we extend this result to higher-order equations.
More precisely, we consider simply typed lambda calculus with product and unit types and study sets of equations between lambda terms (possibly with undefined constants).
Then, for a given set $E$ of equations of lambda terms, we will define a nonnegative integer $e(E)$ that is invariant under equivalence of $E$, i.e., for any $E'$ equivalent to $E$, $e(E) = e(E')$, and show $e(E) \le \#E$ where $\#E$ is the cardinality of set $E$.
Even though $e(E)$ is defined in a homological way, it is algorithmically computable if $E$ is a finite complete \emph{pattern rewriting system (PRS)}, which is a formulation of higher-order rewriting systems introduced in \cite{nipkow1991higher,mayr1998higher}.

\subsubsection*{Historical background}
In many fields of mathematics, homology has been used to bound the size of a given object.
We shall see the case for groups as an example.

In group theory, a \emph{presentation} is a pair of a set $\Sigma$ of \emph{generators} and a set $R\subseteq F(\Sigma)$ of \emph{relations}.
Here, $F(\Sigma)$ is the free group over $\Sigma$,
that is, the set of all words built from elements $a \in \Sigma$ and their formal inverses $a^{-1}$ equipped with multiplication as the word concatenation.
The words $aa^{-1}$ and $a^{-1}a$ are identified with the empty word $1$.
For a presentation $(\Sigma,R)$, the group \emph{presented} by $(\Sigma,R)$ is the group obtained from $F(\Sigma)$ by identifying each $r \in R$ with $1$.

Given a group $G$, one can define an abelian group called the $i$-th \emph{homology} of $G$, written $H_i(G)$, for each $i=0,1,2,\dots$.
Given a group $G$, 
homology of groups has been used in various fields of mathematics, like algebraic topology, number theory, and so on.
Homology is invariant under isomorphism of groups, that is, if $G$ and $G'$ are isomorphic, then $H_i(G)$ and $H_i(G')$ are also isomorphic for all $i$.
One of the well-known facts of homology of groups is the following.
For any group $G$ and any presentation $(\Sigma,R)$ of $G$, we have
\begin{equation}\label{eqn:morseineq}
s(H_2(G))-\rank(H_1(G)) \le \# R - \# \Sigma
\end{equation}
where $s(H)$ is the minimum size of a generating set of $H$, and $\rank(H)$ is the torsion-free rank of $H$,
though the readers have no need to understand their meaning at this stage.
The point here is that we can bound the number of relations using homology.

If we consider monoids instead of groups, a presentation of a monoid is defined as a pair of a set $\Sigma$ of generators and a set $R \subseteq \Sigma^*\times \Sigma^*$ where $\Sigma^*$ is the set of words over $\Sigma$.
We also call such a pair $(\Sigma,R)$ a \emph{string rewriting system (SRS)}.
The monoid presented by $(\Sigma,R)$ is the monoid obtained from $\Sigma^*$ by identifying $l$ with $r$ for each $(l,r) \in R$.
Squier studied monoids using homological algebra in \cite{squier87} and solved an open problem in the field of string rewriting.

The situation for equational axioms is exactly the same.
Function symbols like $\cdot$, ${}^{-1}$, $e$ are considered as generators, and equations are considered as relations.
For first-order equational theories, the algebraic structure \emph{presented by} such generators and relations is a \emph{Lawvere theory} (also called \emph{algebraic theory}) --- a small category with finite products.
Jibladze and Pirashvili studied (co)homology of Lawvere theories in \cite{jibladze1991,jp06}, and the result on bounding the number of equations mentioned earlier is an application of their work.
In more detail, Malbos and Mimram \cite{mm16} showed that the first and second homologies of a Lawvere theory are computable if the theory is presented by a finite and complete term rewriting system.
Then, using their way of computation, Ikebuchi \cite{ikebuchi22} showed that the inequality of form (\ref{eqn:morseineq}) also holds for Lawvere theories.

For higher-order equational theories, it is known that the structure presented by generators (= function symbols) and relations (= higher-order equations) is a \emph{cartesian closed category (CCC)}.
Therefore, we will define the homology of CCCs and use it to show the inequality to bound the number of equations.
For the proof, we introduce the notion of \emph{finite derivation type (FDT)}.
FDT is initially defined for string rewriting systems (SRSs) in \cite{squier94}.
Roughly speaking, FDT is described as follows:
Consider the graph that has strings as vertices and a step of rewriting as an arc.
Then, an SRS is said to have FDT if there is a finite set of cycles, and any cycle in the graph is obtained by combining cycles in the set. 
We extend FDT to higher-order rewriting systems.
%Having FDT is a weaker condition than having a complete PRS, but it is generally more difficult to show FDT than to show completeness.
%We will use FDT for the proof of our main theorem.

\subsubsection*{Outline}
In this paper, we assume the basic knowledge of lambda calculus.
In the next section, we introduce basic terminology and facts on higher-order rewriting.
Then, in Section \ref{sec:mainthm}, we state our main theorem (Theorem \ref{thm:main}) and see some examples.
Our main theorem is written in a form that does not require knowledge of homological algebra.
Section \ref{sec:ccc} provides definitions and facts on CCCs, and introduce the notion of $\Lambda$-sorted CCCs.
In Section \ref{sec:lawvere}, we show that the category of $\Lambda$-sorted CCCs is algebraic and introduce the notion of CCC operations.
In Section \ref{sec:fdt}, we extend the property FDT to CCCs/higher-order equational theories.
Section \ref{sec:ringoid} provides the definitions of ringoids and modules over ringoids.
From this section, we assume some familiarity with basic module theory and homological algebra.
Finally, we prove our main theorem in Section \ref{sec:proof}.

%The first application of homological algebra to rewriting is given in \cite{squier87} for string rewriting.
%A string rewriting system is a pair of an alphabet $\Sigma$ and a subset $R\subset \Sigma^* \times \Sigma^*$ whose elements are called rewrite rules where $\Sigma^*$ is the free monoid.
%Any string $w \in \Sigma^*$ is \emph{rewritten} to $w'$ by $R$, denoted $w \to_R w'$ if there exist $(l,r) \in R$ and $u,v\in\Sigma^*$ such that $w=ulr$ and $w' = urv$.
%Let $\xleftrightarrow{*}_R$ be the reflexive-transitive-symmetric closure of the relation $\to_R$.
%Then, the quotient set $\Sigma^*/\xleftrightarrow{*}_R$ forms a monoid with the multiplication $[w][w'] = [ww']$.
%Conversely, any monoid can be presented by a string rewriting system.

%Before getting into details, let us see why homological algebra can be applied to this problem.
%Recall that the free group generated by an alphabet $\Sigma$ consists of all words built from letters $a$ in $\Sigma$ and their inverses $a^{-1}$.
%For an arbitrary group, 
%Any group can be presented by \emph{generators} and \emph{relations}.

\section{Higher-order rewriting}\label{sec:hrs}

We see basic terminology and facts on higher-order rewriting.
Our formulation of higher-order rewriting is based on \cite{nipkow1991higher,mayr1998higher}.
Those papers treat simply typed lambda calculus, $\lambda^{\to}$, but we add product and unit types to it.
Also, we use the notion of \emph{term-in-context} which suits categorical settings later.

We consider $\lambda^{\to\times1}$, the simply typed lambda calculus with product and unit types.
Let $\Lambda$ be a set of base types.
Our types are defined by the following BNF grammar.
\[
T ::= 1 \text{ (unit)} \mid A (\in \Lambda) \mid T \times T \mid T \to T
\]
We write $\Ty_\Lambda$ for the set of types over $\Lambda$.

A ($\Lambda$-sorted) \emph{signature} is a set of function symbols where each function symbol has a unique type.

We fix a set $V$ of \emph{variables}.
A \emph{typing context} is a finite set of variables with types such that no variable occurs twice or more.
As usual, $x_1:T_1,\dots,x_n:T_n$ denotes the typing context consisting of variables $x_i$ with types $T_i$.

A ($\Lambda$-sorted) \emph{term} $t$ over signature $\Sigma$ is defined by the BNF grammar
\[
t ::= x \mid c \mid () \mid \lambda x:T.~t \mid tt \mid \ab<t,t> \mid \pr_1 t \mid \pr_2 t
\]
where $x$ is a variable in $V$ and $c$ is a function symbol in $\Sigma$.
For a term $t=\lambda x_1:T_1.~\dots.~x_m:T_m.~t_0t_1\dots t_k$, we call $t_0$ the \emph{head} of $t$.

For a typing context $\Gamma$, term $t$, type $T$, the typing relation $\Gamma \vdash t : T$ is defined by the following inference rules.
\[
\begin{bprooftree}
\AxiomC{}
\UnaryInfC{$\Gamma \vdash () : 1$}
\end{bprooftree}
\begin{bprooftree}
\AxiomC{$x:T \in \Gamma \cup \Sigma$}
\UnaryInfC{$\Gamma \vdash x:T$}
\end{bprooftree}
\]
\[
\begin{bprooftree}
\AxiomC{$\Gamma \vdash t:T \to T'$}
\AxiomC{$\Gamma \vdash t' : T$}
\BinaryInfC{$\Gamma \vdash (tt') : T'$}
\end{bprooftree}
\]
\[
\begin{bprooftree}
\AxiomC{$\Gamma,x:T \vdash t:T'$}
\UnaryInfC{$\Gamma \vdash (\lambda x:T.~t) : T'$}
\end{bprooftree}
\begin{bprooftree}
\AxiomC{$\Gamma \vdash t_1:T_1$}
\AxiomC{$\Gamma \vdash t_2 : T_2$}
\BinaryInfC{$\Gamma \vdash \ab<t_1,t_2> : T_1\times T_2$}
\end{bprooftree}
\]
\[
\begin{bprooftree}
\AxiomC{$\Gamma \vdash t : T_1\times T_2$}
\UnaryInfC{$\Gamma \vdash \pr_1t : T_1$}
\end{bprooftree}
\begin{bprooftree}
\AxiomC{$\Gamma \vdash t : T_1\times T_2$}
\UnaryInfC{$\Gamma \vdash \pr_2t : T_2$}
\end{bprooftree}
\]
If $\Gamma \vdash t : T$ holds, we call the expression $\Gamma \vdash t : T$ (or just $\Gamma \vdash t$) a \emph{term-in-context} or call $t$ a \emph{term in context} $\Gamma$.

For typing contexts $\Gamma,\Gamma'$, a $\Gamma$-$\Gamma'$-\emph{substitution} is a mapping from $x:T \in \Gamma$ to a term-in-context $\Gamma' \vdash t:T$.
A $\Gamma$-$\Gamma'$-substitution $\theta$ can be extended to a function that maps $\Gamma\vdash t:T$ to $\Gamma' \vdash t':T$ where $t'$ is the term obtained from $t$ by replacing each occurrence of variable $x$ in $\Gamma$ with $\theta(x)$.
We write $t\theta$ for such $t'$.

For a relation $\sim$ defined between two terms with the same type in the same context, we write $\Gamma \vdash t \sim t' : T$ or $\Gamma \vdash t \sim t'$ instead of $(\Gamma \vdash t : T) \sim (\Gamma \vdash t' : T)$.

A relation $\Gamma \vdash t \sim t'$ is \emph{compatible} if it satisfies the following conditions.
\begin{itemize}
  \item If $\Gamma \vdash t \sim t'$, then $\Gamma' \vdash t\theta \sim t'\theta$ for any $\Gamma$-$\Gamma'$-substitution $\theta$.
  \item If $\Gamma \vdash t_1 \sim t_1' : T \to T'$ and $\Gamma \vdash t_2 \sim t_2' : T$, then $\Gamma \vdash t_1t_2 \sim t_1't_2' : T'$.
  \item If $\Gamma, x:T \vdash t \sim t'$, then $\Gamma \vdash (\lambda x:T.~t) \sim (\lambda x:T.~t')$.
  \item If $\Gamma \vdash t_1 \sim t_1'$ and $\Gamma \vdash t_2 \sim t_2'$, then $\Gamma \vdash \ab<t_1,t_2> \sim \ab<t_1',t_2'>$.
  \item If $\Gamma \vdash t \sim t'$, then $\Gamma \vdash \pr_i t \sim \pr_i t'$ for $i=1,2$.
\end{itemize} 
A compatible relation is a \emph{congruence} if it is an equivalence relation.

The notion of $\beta$-reduction is defined as usual.
We define the $\delta$-reduction rule $\to_\delta$ as the compatible closure of
\[
\pr_i(\ab<t_1,t_2>) \to_\delta t_i \quad (i=1,2).
\]
A term $t$ is said to be $\beta\delta$-\emph{normal} if $\beta$- and $\delta$-reductions cannot be applied to $t$.

We define the $\eta$-expansion rule $\to_{\bar\eta}$ on term-in-contexts as the compatible closure of the following rules.
\[
\Gamma \vdash t \to_{\bar\eta} (\lambda x:T.~t x) : T \to T',
\]
\[
\Gamma \vdash t \to_{\bar\eta} \ab<\pr_1 t, \pr_2 t> : T \times T'.
\]
For a $\beta\delta$-normal term $t$, the term-in-context $\Gamma\vdash t$ is said to be $\beta\delta\bar\eta$-normal if it satisfies the following:
If $\Gamma \vdash t \to_{\bar\eta} t'$, then $t'$ is not $\beta\delta$-normal.

The $\beta\delta$-, $\eta$-, $\beta\delta\eta$-equivalences are the smallest congruences containing $\to_\beta\cup \to_\delta$, $\to_{\bar\eta}$, $\to_{\bar\eta}\cup \to_\beta\cup \to_\delta$, respectively.
%For a $\beta\delta$-normal term $t$, we define the $\eta$-\emph{expanded form} of $t$, denoted $e\uparrow^\eta$ as follows.
%\begin{align*}
%  &\ab(\lambda x_1\dots x_n .t t_1\dots t_m)\uparrow^\eta\\ &= \lambda x_1\dots x_n x_{n+1}\dots x_{n+k}. t \ab(t_1\uparrow^\eta)  \dots \ab(t_m\uparrow^\eta) x_{n+1}\dots x_{n+k}
%\end{align*}
%for $t : T_1\times \dots T_{n+k} \to T$, and
%\begin{align*}
%  \ab< t_1,t_2 >\uparrow^\eta &= \ab< \ab(t_1\uparrow^\eta), \ab(t_2\uparrow^\eta) >,\\
%  \ab(\pr_i t)\uparrow^\eta &= \pr_i(t\uparrow^\eta) \quad (i=1,2).
%\end{align*}

%We can show that $\to_{\beta}\cup\to_{\delta}$, 
%For a term $t$, we write $e\downarrow_{\beta\delta}$ for the $\beta\delta$-normal form of $t$ and $t\updownarrow_{\beta\delta}^{\eta}$ for the $\beta\delta\bar\eta$-normal form of $t$.
%We call $t\updownarrow_{\beta\delta}^\eta$ the $\beta\delta\bar\eta$-normal form of $t$.
For $\beta\delta$-normal terms $t,t_1,t_2$, let $t[t_1/t_2]$ denote the term obtained from $t$ by replacing each subterm $t_2$ with $t_1$.

We define the notions of free and bound variables, unification, most general unifiers for terms in the usual way.
%For term-in-context $\Gamma \vdash t$, we say that a variable $x$ is \emph{free} (resp. \emph{bound}) in $\Gamma \vdash t$ if $x$ is a free (resp. bound) variable in $t$.

Consider a transformation of term-in-context of form $\Gamma, x : T_1 \times T_2 \vdash t$ to $\Gamma, x_1 : T_1,x_2 : T_2 \vdash t'$ where $x_1,x_2$ are fresh variables and $t'$ is the term obtained from $t$ by replacing each occurrence of $x$ with $\ab<x_1,x_2>$ and apply $\delta$-reductions if possible.
Repeating this, we obtain $\Gamma'' \vdash t''$ where for each $x_o:T_i \in \Gamma''$, $T_i$ is not a product type.
We call such $\Gamma'' \vdash t''$ the \emph{depaired form} of $\Gamma \vdash t$.

A term-in-context $\Gamma \vdash t$ in $\beta\delta$-normal form is a \emph{pattern} if its depaired form is $\Gamma'\vdash t'$ and every free variable $x$ in $t'$ occurs in the form $xx_1\dots x_n$ where $x_1,\dots,x_n$ are $\eta$-equivalent to distinct bound variables.
A reason to consider patterns is that whether two patterns are unifiable is decidable \cite{miller1991,nipkow1993}.
Those papers consider $\lambda^\to$, but it is not difficult to extend that result to $\lambda^{\to\times1}$.
For example, $x:(T \to T) \to T \vdash \lambda y:(T\to T).~c(x(\lambda z:T.~yz))$ (for $c:T\to T \in \Sigma$) and $x:(T \to T \to T)\times T \vdash \lambda y:T.~\lambda z:T.~(\pr_1 x)zy$ are patterns but $x:T \to T \vdash xc$ (for $c : T \in \Sigma$) and $x:T \to T \to T \vdash \lambda y:T.~xyy$ are not patterns.

A pair of two $\beta\delta$-normal term-in-contexts $\Gamma \vdash t:T$ and $\Gamma' \vdash t':T'$ is an \emph{equation-in-context} if $\Gamma = \Gamma'$, $T=T'$ and $T$ is a base type.
An equation-in-context is written as $\Gamma \vdash t \approx t' : T$ or $\Gamma \vdash t \approx t'$.
We also say that $t \approx t'$ is an equation in context $\Gamma$.
We call a set of equation-in-contexts an \emph{equation system}.
Also, we call the pair $(\Sigma,E)$ of a signature and an equation system an \emph{equational theory}.

An equation-in-context $\Gamma \vdash t\approx t'$ is a \emph{rule-in-context} if the free variables in $t'$ appear in $t$ as free variables.
A rule-in-context is written as $\Gamma \vdash t \to t' : T$ or $\Gamma \vdash t \to t'$.
We also say that $t \to t'$ is a rule in context $\Gamma$.
A \emph{higher-order rewriting system (HRS)} is a set of rule-in-contexts.

An HRS $R$ is a \emph{pattern rewriting system (PRS)} if the left-hand side of each rule-in-context in $R$ is a pattern.

For an equation system $E$, we define a relation $\Gamma \vdash t \approx_E t'$ as the smallest congruence satisfying (i) $\Gamma \vdash t \approx_E t'$ if $t$ and $t'$ are $\beta\delta\eta$-equivalent, and (ii) $\Gamma \vdash t \approx_E t'$ if $\Gamma \vdash t \approx t' : T$ is in $E$.

Two equation systems $E,E'$ are \emph{equivalent} if $\Gamma \vdash t \approx_E t' \Leftrightarrow \Gamma \vdash t \approx_{E'} t'$ for any $\Gamma,t,t'$. 

For a term $t$, any subterm of $t$ can be specified by a \emph{position}.
Formally, a position is a list $\fp = n_1\dots n_k$ of positive integers.
We write $[]$ for the empty list.

For a term $t$ and a position $\fp$, the subterm of $t$ at $\fp$, denoted by $t|_\fp$, is defined as follows:
\begin{align*}
  t|_{[]} &= t,
  &(t_1t_2)|_{i\fp} &= t_i|_\fp,\\
  (\lambda x:T.~t)|_{1\fp} &= t|_\fp,
  &(\ab<t_1,t_2>)|_{i\fp} &= t_i|_\fp,\\
  (\pr_i t)|_{1\fp} &= t|_\fp,
\end{align*}
where $i=1,2$.

Also, we write $e[u]_\fp$ for the term obtained from $e$ by replacing the subterm at $\fp$ with $u$.

We define an order $\succ$ of positions as follows: $\fp_1 \succ \fp_2$ if and only if $\fp_2$ is an prefix of $\fp_1$.
%We say that $\fp_1$ is \emph{below} $\fp_2$ if $\fp_1 \succ \fp_2$.
We say that two positions $\fp_1$, $\fp_2$ are \emph{disjoint} if they are incomparable with respect to $\succ$.

For a rule-in-context $\Gamma \vdash l \to r$, the relation $\Gamma' \vdash t \xrightarrow[\Gamma' \vdash l \to r]{} t'$ is defined as follows:
$\Gamma' \vdash t \xrightarrow[\Gamma' \vdash l \to r]{} t'$ holds if and only if there exist a position $\fp$ and a $\Gamma$-$\Gamma'$ substitution $\theta$ such that $t|_\fp = l\theta$ and $t' = t[r\theta]_\fp$.

The relation $\Gamma \vdash t \to_R' t'$ holds if and only if $\Gamma \vdash t \xrightarrow[\Gamma' \vdash l \to r]{} t'$ for some $\Gamma' \vdash l \to r$ in $R$.
Also, define the relation $\to_R$ as follows: $\Gamma \vdash t \to_R t'$ if and only if  $\Gamma \vdash \ab(t\updownarrow_{\beta\delta}^\eta) \to_R' \ab(t'\updownarrow_{\beta\delta}^\eta)$
where $t\updownarrow_{\beta\delta}^\eta$ is the $\beta\delta\bar\eta$-normal form of $t$.
We write $\to_R^*$ for the reflexive transitive closure of $\to_R$.
We can show that the symmetric closure of $\to_R^*$ coincides with $\approx_R$.

A term-in-context $\Gamma \vdash t$ is $R$-\emph{normal} if there is no term-in-context $\Gamma' \vdash t'$ such that $\Gamma \vdash t \to_R t'$.

We say that $R$ is \emph{terminating} if $\to_R$ is well-founded.
We say that $R$ is \emph{confluent} if $\Gamma \vdash t \to_R^* t_1$ and $\Gamma \vdash t \to_R^* t_2$ imply that there exists $t'$ such that $\Gamma \vdash t_1 \to_R^* t'$ and $\Gamma \vdash t_2 \to_R^* t'$.
We say that $R$ is \emph{complete} if $R$ is both confluent and terminating.

For first-order term rewriting systems, to prove a terminating system is confluent, it suffices to check if each \emph{critical pair} can be rewritten to the same term.
In \cite{nipkow1991higher,mayr1998higher}, critical pairs are extended to PRSs.
Again, these papers consider $\lambda^\to$, but we can define critical pairs for $\lambda^{\to\times1}$ in the same way as follows.

Assume that $\Gamma_i \vdash l_i \to r_i$ ($i=1,2$) are two rule-in-contexts such that their left- and right-hand sides are depaired. If they are not depaired, we take their depaired form.
We say that $\Gamma_1\vdash l_1 \to r_1$ and $\Gamma_2 \vdash l_2 \to r_2$ \emph{overlaps at} $\fp$ if 
\begin{itemize}
  \item the head of $l_1|_\fp$ is not a free variable in $l_1$, and
  \item the two patterns $\lambda x_1:T_1\dots x_k:T_k.~l_1|_\fp$ and $\lambda x_1:T_1\dots x_k:T_k.~l_2$ have a most general unifier $\theta$ where $x_1,\dots,x_k$ are the variables that are bound in $l_1$ but free in $l_1|_\fp$ and $T_i$ is the type of $x_i$ for each $i$.
\end{itemize}
Here, we rename the variables so that the free variables in $l_1$, the bound variables in $l_1$, and the free variables in $l_2$ are all distinct.

We call the diagram $r_1\theta \leftarrow l_1\theta = (l_1[l_2]_\fp)\theta \to (l_1[r_2]_\fp)\theta$ a \emph{critical peak} and the pair $(r_1\theta, (l_1[r_2]_\fp)\theta)$ a \emph{critical pair} between $\Gamma_1 \vdash l_1 \to r_1$ and $\Gamma_2 \vdash l_2 \to r_2$.
The term $l_1\theta = (l_1[l_2]_\fp)\theta$ is called the \emph{superposition}.

%The following lemma ensure that critical pairs defined as above is 
%\begin{lemma}[Critical Pair Lemma]
%Let $R$ be a PRS.
%If $t$ is rewritten to $t_i$ by $\Gamma_i\vdash l_i \to r_i \in R$ at $p_i$ ($i=1,2$), then either (i) $t_1 \to_R^* \hat t \leftarrow_R^* t_2$ for some $\hat t$ or (ii) there exist terms $t'_1,t'_2$, substitution $\theta$, and a position $\fp$ such that $t_i = t[t'_i\theta]_\fp$ ($p=1,2$) and $(t_1',t_2')$ is a critical pair.
%\end{lemma}
%The proof is similar to the one given in \cite{mayr1998higher} for simply typed lambda calculus without product types.

\section{Main Theorem}\label{sec:mainthm}

Let $\Sigma$ be a $\Lambda$-sorted signature and $R$ be a complete PRS such that for any rule $\Gamma \vdash l \to r$ in $R$, the multisets of free variables in $l$ and $r$ are the same.
This assumption on free variables is crucial for the proof of our main theorem, especially for Definition \ref{def:coefmulti}.
Suppose that $R$ has $n$ rule-in-contexts $\Gamma_i \vdash l_i \to r_i$ ($i=1,\dots,n$) and $k$ critical pairs $(t_j,s_j)$ between $\Gamma_{a_j} \vdash l_{a_j} \to r_{a_j}$ and $\Gamma_{b_j} \vdash l_{b_j} \to r_{b_j}$ ($a_j,b_j \in \ab\{1,\dots,n\}$) ($j=1,\dots,k$). 

For each term-in-context $\Gamma \vdash t$, we choose a path of rewriting $\Gamma \vdash t \to_R \dots \to_R \hat t$ from $\Gamma \vdash t$ to the $R$-normal form $\Gamma\vdash \hat t$.
\begin{definition}
The \emph{second boundary matrix} $D_2(R)$ is the $n\times k$ matrix whose $(i,j)$-th entry is given as follows:
Take any fixed chosen rewriting paths $t_j \to_R \dots \to_R \hat t_j$ and $s_j \to_R \dots \to_R \hat s_j (= \hat t_j$) from $t_j$, $s_j$ to their normal form $\hat t_j = \hat s_j$.
Let $N_{i,j}$ (resp. $M_{i,j}$) be the number of times $\Gamma_i\vdash l_i \to r_i$ is used in the path $u_j \xrightarrow[\Gamma_{a_j} \vdash l_{a_j}\to r_{a_j}]{} t_j \to_R \dots \to_R \hat t_j$ (resp. $u_j \xrightarrow[\Gamma_{b_j}\vdash l_{b_j}\to r_{b_j}]{} s_j \to_R \dots \to_R \hat s_j$). 
Then, the $(i,j)$-th entry of $D_2(R)$ is defined as $N_{i,j} - M_{i,j}$.
\end{definition}

%\begin{definition}[Smith normal form]
%A $k\times n$ matrix $M$ over $\bbQ$ is in \emph{Smith normal form} if only $(i,i)$th entries $M_{i,i}$ of $M$ is non-zero for $i\in \ab\{1,\dots,\min(k,n)\}$ and $M_{i,i}$ divides $M_{i+1,i+1}$ for $i \in \ab\{1,\dots,\min(k,n)-1\}$.
%\end{definition}
%It is well-known that any matrix over $\bbZ$ can be transformed into Smith normal form by elementary row/column operations, that is, (i) switching a row/column with another row/column, (ii) adding a multiple of a row/column to another row/column, and (iii) multiplying a row/column by -1.
%
%\begin{definition}
%For a matrix $M$, the integer $e(M)$ is the number of $\pm 1$s in a Smith normal form of $M$.
%\end{definition}
The following is our main theorem.
\begin{theorem}\label{thm:main}
For any equation system $E$ equivalent to $R$, the following inequality holds:
\[
\# R - \rank(D_2(R)) \le \#E.
\]
Here, $\rank(D_2(R))$ is the rank of $D_2(R)$ as a matrix over the field of rationals $\bbQ$.
\end{theorem}

We shall see a simple example.
Let $\Lambda_2 = \{U,V\}$, $\Sigma_2 = \{\lnot : U \to U, {\land},{\lor} : U \to U \to U, \forall, \exists : (V \to U) \to U\}$.
Consider PRS $R_2$ consisting of
\begin{align}
  x:U &\vdash \lnot \lnot x \to x,\tag{\rm NotNot}\label{eqn:notnot}\\
  x:U, y:U &\vdash \lnot (x \land y) \to \lnot x \lor \lnot y, \tag{\rm NotAnd}\label{eqn:notand}\\
  x:U, y:U &\vdash \lnot (x \lor y) \to \lnot x \land \lnot y, \tag{\rm NotOr}\label{eqn:notor}\\
  p:V \to U &\vdash \lnot \forall (\lambda z:V.~pz) \to \exists (\lambda z:V.~\lnot pz), \tag{\rm NotAll}\label{eqn:notall}\\
  p:V \to U &\vdash \lnot \exists (\lambda z:V.~pz) \to \forall (\lambda z:V.~\lnot pz) \tag{\rm NotEx}\label{eqn:notex}.
\end{align}
This PRS $P_2$ rewrites a formula to its negation normal form.
Again, we can prove that $R_2$ is complete \cite{mayr1998higher} and has 5 critical peaks as described in Figure \ref{fig:cps2}.
Here, for example, the edge label (\ref{eqn:notor})(\ref{eqn:notnot})(\ref{eqn:notnot}) in $\Pi_2'$ means that we need (\ref{eqn:notor}) once and (\ref{eqn:notnot}) twice to normalize $\lnot(\lnot x \lor \lnot y)$ into $x \land y$.
The matrix $D_2(R_2)$ is given as
\[
\bordermatrix{
  & \Pi_1' & \Pi_2' & \Pi_3' & \Pi_4' & \Pi_5'\cr
 (\ref{eqn:notnot})& 0 & 1 & 1 & 0 & 0\cr
 (\ref{eqn:notand})& 0 & 1 & 1 & 0 & 0\cr
 (\ref{eqn:notor})& 0 & 1 & 1 & 0 & 0\cr
 (\ref{eqn:notall})& 0 & 0 & 0 & 1 & 1\cr
 (\ref{eqn:notex})& 0 & 0 & 0 & 1 & 1\cr
}
\]
and get $\rank D_2(R_2) = 2$.
So, any equation system equivalent to $R_2$ has at least 3 ($= 5-2$) equations.
In fact, the system $\{(\ref{eqn:notnot}),(\ref{eqn:notor}),(\ref{eqn:notex})\}$ is equivalent to $R_2$ because (\ref{eqn:notand}) can be derived as
\begin{align*}
\lnot (x \land y) \xleftarrow[(\ref{eqn:notnot})]{*} \lnot (\lnot \lnot x \land \lnot \lnot y) \xleftarrow[(\ref{eqn:notor})]{} \lnot \lnot (\lnot x \lor \lnot y),\\
 \lnot \lnot (\lnot x \lor \lnot y)\xrightarrow[(\ref{eqn:notnot})]{} \lnot x \lor \lnot y
\end{align*}
and (\ref{eqn:notall}) can be derived similarly.

\begin{figure*}
\begin{tikzcd}
(\Pi_1')&\\
\lnot (\lnot x) \ar[r, shift left, "(\ref{eqn:notnot})"] \ar[r, shift right, "(\ref{eqn:notor})"']& x
\end{tikzcd}
\begin{tikzcd}
(\Pi_2') & \lnot (\lnot x \lor \lnot y) \ar[d, "(\ref{eqn:notor})(\ref{eqn:notnot})(\ref{eqn:notnot})"]\\
\lnot (\lnot (x \land y)) \ar[r, "(\ref{eqn:notnot})"'] \ar[ur, shift right, "(\ref{eqn:notand})"] & x \land y
\end{tikzcd}
\begin{tikzcd}
(\Pi_3')& \lnot (\lnot x \land \lnot y) \ar[d, "(\ref{eqn:notand})(\ref{eqn:notnot})(\ref{eqn:notnot})"]\\
\lnot (\lnot (x \lor y)) \ar[r, "(\ref{eqn:notnot})"'] \ar[ru, shift right, "(\ref{eqn:notor})"] & x \lor y
\end{tikzcd}
\begin{tikzcd}
(\Pi_4') & \lnot \exists (\lambda z:V.~\lnot pz) \ar[d, "(\ref{eqn:notex})(\ref{eqn:notnot})"]\\
\lnot (\lnot \forall (\lambda z:V.~pz)) \ar[r, "(\ref{eqn:notnot})"'] \ar[ru, shift right, "(\ref{eqn:notall})"]& \forall (\lambda z:V.~pz)
\end{tikzcd}
\begin{tikzcd}
(\Pi_5') & \lnot \forall (\lambda z:V.~\lnot pz) \ar[d, "(\ref{eqn:notall})(\ref{eqn:notnot})"]\\
\lnot (\lnot \exists (\lambda z:V.~pz)) \ar[r, "(\ref{eqn:notnot})"'] \ar[ru, shift right, "(\ref{eqn:notex})"]& \exists (\lambda z:V.~pz)
\end{tikzcd}
\caption{\label{fig:cps2} Critical peaks of $R_2$ and their normalization.}
\end{figure*}

\section{Cartesian closed categories}\label{sec:ccc}

In this section, we review the notion of cartesian closed categories.
For more details, see \cite{Crole1994} for example.

\begin{definition}
Let $\bfC$ be a category.
Given objects $X_1,\dots,X_n$ of $\bfC$, a \emph{product} of them is an object $X_1\times \dots\times X_n$ of $\bfC$ together with morphisms $\pi_i : X_1\times\dots\times X_n \to X_i$ called \emph{$i$-th projection} ($i=1,\dots,n$) such that for any object $Z$ and morphisms $f_i : Z \to X_i$, there exists a unique morphism $\ab<f_1,\dots,f_n> : Z \to X_1\times \dots \times X_n$ such that $f_i = \pi_i \circ \ab<f_1,\dots,f_n>$ for each $i\in I$.

We say that $\bfC$ has \emph{finite products} if it has a product of $X_1,\dots,X_n$ for any natural number $n$.
\end{definition}

A \emph{coproduct} is defined as the dual of a product.
\begin{definition}
Let $\bfC$ be a category.
Given objects $X_1,\dots,X_n$ of $\bfC$, a \emph{coproduct} of them is an object $X_1+\dots+X_n$ of $\bfC$ together with morphisms $\iota_i : X_i \to X_1+\dots+X_n$ ($i=1,\dots,n$) such that for any object $Z$ and morphisms $f_i : X_i \to Z$, there exists a unique morphism $\ab[f_1,\dots,f_n] : X_1+\dots+X_n \to Z$ such that $f_i = [f_1,\dots,f_n] \circ \iota_i$ for each $i\in I$.
\end{definition}

\begin{definition}
Let $\bfC,\bfD$ be two categories having finite products.
A functor $F : \bfC \to \bfD$ is said to \emph{(strictly) preserve finite products} if for any family $\{X_1,\dots,X_n\}$ of objects of $\bfC$ and projections $\pi_i : X_1\times\dots\times X_n \to X_i$ ($i=1,\dots,n$),
%the morphism
$$\ab< F\pi_1,\dots,F\pi_n > : F(X_1\times\dots\times X_n) \to FX_1\times \dots \times FX_n$$
is an identity in $\bfD$.
In particular,
$$F(X_1\times\dots\times X_n) = FX_1\times \dots \times FX_n.$$
\end{definition}

\begin{definition}
Let $\bfC$ be a category.
For objects $Y,Z$ of $\bfC$, an object $Z^Y$ together with a morphism $\ev : Z^Y \times Y \to Z$ is an \emph{exponential} if for any $X$ and $g : X\times Y \to Z$, there exists a unique $\lambda g : X \to Z^Y$ such that $\ev \circ (\lambda g \times \Id_Y) = g$.

We say that $\bfC$ is a \emph{cartesian closed category (CCC)} if it has all finite products and exponentials of any two objects.
\end{definition}

\begin{definition}
For two CCCs $\bfC$, $\bfD$, a functor $F : \bfC \to \bfD$ that preserves finite products is said to \emph{(strictly) preserve exponentials} if for any $X,Y$ in $\bfC$, $F(Y^X) = FY^{FX}$ and $\lambda(F\ev) = \Id_{FY^{FX}}$.
%\[
%F(X\Rightarrow Y)\times FX = F((X\Rightarrow Y)\times X) \xrightarrow{F\ev} FY.
%\]
A functor $F : \bfC \to \bfD$ is a \emph{(strict) cartesian closed functor} if $F$ preserves finite products and exponentials.
\end{definition}

Let $\Lambda$ be a set, $\Sigma$ an $\Lambda$-sorted signature,
and $E$ an equation system over $\Sigma$. 
For two term-in-contexts $x : T \vdash t_1 : T'$ and $y : T \vdash t_2 : T'$, we define an equivalence relation $\sim$ by
\[
(x : T \vdash t_1 : T') \sim (y : T \vdash t_2 : T') \text{ iff } x:T \vdash t \approx_E t'[x/y].
\]
We write $(x : T \mid_E t : T')$ for the equivalence class of $x : T \vdash t : T'$ with respect to $\sim$.

We define the category $\Cl(\Sigma,E)$, called the \emph{canonical classifying category} or the \emph{CCC presented by} $(\Sigma,E)$, as follows.
\begin{itemize}
  \item An object is a type in $\Ty_\Lambda$.
  \item A morphism from $T$ to $T'$ is $(x : T \mid_{E} t : T')$.
  \item Composition $(y : T' \mid_E t' : T'') \circ (x : T \mid_E t : T')$ is
\[
(x : T \mid_E t'[t/y] : T'').
\]
  \item The identity on type $T$ is $(x:T \mid_E x : T)$.
\end{itemize}
We can check that $\Cl(\Sigma,E)$ is cartesian closed.
For example, $\ev : Z^Y \times Y \to Z$ is given as
\[
\ab(z:Z^Y\times Y \mid_E (\pr_1 z)(\pr_2 z) )
\]
and, for $\tau = (w:X\times Y \mid t:Z)$, $\lambda \tau$ is given as
\[
\ab(x:X \mid_E \lambda y:Y.~t[\ab<x,y>/w]).
\]

We write $\Cl(\Sigma)$ for $\Cl(\Sigma,\emptyset)$ and call it the \emph{CCC freely generated by $\Sigma$}.
We sometimes write $\Cl(\{T_1 \to T_1',\dots,T_n \to T_n'\})$ instead of $\Cl(\{c_1 : T_1 \to T_1',\dots,c_n : T_n \to T_n'\})$.
Also, we write $\bar c_i$ for the morphism $(x:T_{i} \mid_\emptyset c_ix : T_{i}')$ in $\Cl(\{c_1:T_1 \to T_1',\dots, c_n : T_n \to T_n'\})$.

Note that any morphism $(x:T\mid_\emptyset t:T')$ in $\Cl(\Sigma)$ is the $\alpha\beta\delta\eta$-equivalence class of term-in-context $x:T \vdash t : T'$.

\subsection{Cartesian closed categories with a fixed set of sorts}
In this subsection, we introduce a notion of $\Lambda$-\emph{sorted CCCs}.
Fix a set $\Lambda$.
We write $\catcfam_\Lambda$ for the CCC $\Cl(\emptyset)$ where $\emptyset$ is considered as the empty $\Lambda$-sorted signature.

\begin{definition}
A $\Lambda$-\emph{sorted CCC} is a CCC $\bfC$ together with a (strict) cartesian closed functor $\iota \colon \catcfam_\Lambda \to \bfC$ such that $\Ob(\bfC) = \Ty_\Lambda$ and $\iota$ is identity on objects.
We often call $\bfC$ a $\Lambda$-sorted CCC without mentioning $\iota$.

A \emph{morphism} from a $\Lambda$-sorted CCC $\iota:\catfam_\Lambda \to \bfC$ to another $\Lambda$-sorted CCC $\iota' : \catfam_\Lambda \to \bfC$ is a cartesian closed functor $F : \bfC \to \bfC'$ such that $F\circ\iota = \iota'$.

$\Lambda$-sorted CCCs and morphisms between them form a category and let  $\catccc_\Lambda$ denote the category.
\end{definition}
The following states that any CCC has a structure of $\Lambda$-sorted CCC for some $\Lambda$.
\begin{proposition}
For any CCC $\bfC$, there exists a set $\Lambda$ and an $\Lambda$-sorted CCC $\catcfam_\Lambda \to \tilde\bfC$ such that $\bfC$ is equivalent to $\tilde\bfC$.
\end{proposition}
\begin{IEEEproof}
Let $\Lambda = \Ob(\bfC)$. Define $\tilde\bfC$ by $\Hom_{\tilde\bfC}(X,Y) = \Hom_\bfC(FX,FY)$ where $FX$ for $X \in \Ty_\Lambda$ is defined as (i) $FX = 1$ if $X = 1$, (ii) $FX = X$ if $X \in \Lambda$, (iii) $FX = FY \times FZ$ if $X = Y\times Z$, and (iv) $FX = FZ^{FY}$ if $X = (Y \to Z)$.
Then, we have a full and faithful functor $\tilde\bfC \to \bfC$ that is surjective on objects.
\end{IEEEproof}

For any $\Lambda$-sorted signature $\Sigma$ and an equation system $E$ over $\Sigma$, $\Cl(\Sigma,E)$ is a $\Lambda$-sorted CCC.

%\section{Universal algebra}
%It is not very difficult to see that the category of $\Lambda$-sorted CCCs is algebraic, i.e., it is equivalent to the category of models of a (many-sorted) first-order equational theory.
%
%Let $S = \Ty_\Lambda \times \Ty_\Lambda$ and construct an $S$-sorted first-order equational theory.
%Let $\Sigma$ consist of function symbols $\mathsf{id}_T : 1 \to (T,T)$, $\circ_{T_1,T_2,T_3} : (T_2,T_3)\times(T_1,T_2) \to (T_1,T_3)$, $!_T : 1 \to (T, 1)$, $\ab<\_,\_> : (T,T_1)\times(T,T_2) \to (T,T_1\times T_2)$, $\pr_i^{T_1,T_2} : 1 \to (T_1\times T_2,T_i)$ ($i=1,2$), $\ev^{T_1,T_2} : 1 \to ((T_1 \to T_2)\times T_1,T_2)$, and $\lambda^{T,T_1,T_2} : ((T\times T_1,T_2) \to (T,T_1 \to T_2))$ for each $T,T',T_1,T_2,T_3 \in \Ty_\Lambda$.
%We often omit the subscripts or superscripts.

\section{Lawvere theory of $\Lambda$-sorted CCCs}\label{sec:lawvere}
In this section, we consider a Lawvere theory of $\Lambda$-sorted CCCs.

A \emph{(many-sorted) Lawvere theory} is a small category with finite products.
For a Lawvere theory $\bfT$, a $\bfT$-\emph{model} is a functor $M : \bfT \to \catset$ that preserves finite products.
%If $M$ is a $\bfT$-model and $s : X_1\times \dots \times X_n \to X$ is a morphism in $\bfT$, we write $\llbracket s \rrbracket_M$ for the map $Ms : MX_1\times \dots \times MX_n \to MX$.

A \emph{morphism} between two $\bfT$-models $M,N$ is a natural transformation $\alpha : M \to N$.
For a Lawvere theory $\bfT$, the collection of $\bfT$-models and morphisms between them forms a category denoted by $\Alg\bfT$.
It is well-known that any Lawvere theory $\bfT$ can be presented by a set of first-order equations and $\Alg\bfT$ is equivalent to the category of models of the set of equations \cite{lawvere1963}.

Let $(\catccc_\Lambda)_\pp$ be the full subcategory of $\catccc_\Lambda$ consisting of objects $\Cl(\Sigma)$ for $\Lambda$-sorted finite signatures $\Sigma$.

We can check that $(\catccc_\Lambda)_\pp$ has binary coproducts as $$\Cl(\Sigma)+ \Cl(\Sigma') = \Cl(\Sigma\uplus\Sigma')$$
and initial object $\Cl(\emptyset)$.
So, $(\catccc_\Lambda)_\pp^\Op$ is a Lawvere theory.
\begin{proposition}
We have an equivalence of categories $$\Alg\ab((\catccc_\Lambda)_\pp^\Op) \simeq \catccc_\Lambda.$$
\end{proposition}
\begin{IEEEproof}
For a $\Lambda$-sorted CCC $\bfC$, let $M_\bfC : (\catccc_\Lambda)_\pp^\Op \to \catset$ be the functor that maps objects as
\[
M_\bfC(\Cl(\{c:T \to T'\})) =  \Hom_{\bfC}(T,T')
\]
and maps a morphism $F$ from $\Cl(\{c:T \to T'\})$ to $\Cl(\{c_1:T_1 \to T_1',\dots,c_n:T_n \to T_n'\})$ of $\Lambda$-sorted CCCs as follows.
Given $f_1:T_1 \to T_1', \dots, f_n : T_n \to T_n'$ in $\bfC$, let $G : \Cl(\{c_1:T_1 \to T_1',\dots,c_n:T_n \to T_n'\}) \to \bfC$ be the cartesian closed functor that maps $\bar c_i$ to $f_i$.
Then,
\(
M_\bfC (F) : \Hom_\bfC(T_1,T_1')\times \dots\times \Hom_\bfC(T_n,T_n') \to \Hom_\bfC(T,T')\),
\(
M_\bfC (F)(f_1,\dots,f_n) = GF(c)\).

Conversely, for a model $M : (\catccc_\Lambda)_\pp^\Op \to \catset$, we can define a $\Lambda$-sorted CCC $\bfC_M$ as follows.
$\Hom_{\bfC_M}(T,T') = M(\Cl(\{c:T \to T'\}))$ for a fresh symbol $c$ and, for $f\in\Hom_{\bfC_M}(T,T')$ and $f'\in\Hom_{\bfC_M}(T',T'')$, $f' \circ f$ is defined as $M(F^\Op)(f,f')$ where $F : \Cl(\{c : T \to T''\}) \to \Cl(\{c_1 : T\to T', c_2 : T' \to T''\})$ is the cartesian closed functor that maps $\bar c$ to $\bar c_2 \circ \bar c_1$.

Also, for a morphism between models $\alpha : M \to N$, we can define a cartesian closed functor $F_\alpha : \bfC_M \to \bfC_N$ as $F_\alpha(f) = \alpha_{\Cl(\{T \to T'\})}(f)$ for $f : T \to T'$.
\end{IEEEproof} 

Note that a Lawvere theory for $\Lambda$-sorted CCCs can be constructed in different ways. For example, \emph{categorical combinators} \cite{curien86} explicitly give a first-order equational theory for $\Lambda$-sorted CCCs.

\begin{definition}
A \emph{CCC operation} is a morphism $\omega : \Cl(\{T \to T'\}) \to \Cl(\{T_1 \to T_1',\dots,T_n \to T_n'\})$ of $\Lambda$-sorted CCCs such that $T',T_1',\dots,T_n'$ are base types.
\end{definition}
Note that any CCC operation $\omega : \Cl(\{c:T \to T'\}) \to \Cl(\{\square_1:T_1 \to T_1',\dots,\square_n:T_n \to T_n'\})$ is uniquely determined by $\omega(\bar c)$.
We write $\llparenthesis \tau \rrparenthesis$ for the CCC operation determined by $\tau$.

\begin{definition}[multiplicity $\multiplicity{c_i}{\omega}$]
Let $\omega : \Cl(\{c:T \to T'\}) \to \Cl(\{c_1:T_1 \to T_1',\dots,c_n:T_n \to T_n'\})$ be a CCC operation.
Suppose that $\omega$ is determined by $(x:T \mid_\emptyset t:T')$ and $t$ is $\beta\delta\bar\eta$-normal.
The \emph{multiplicity} of $\omega$ in $c_i$, denoted $\multiplicity{c_i}{\omega}$, is the number of occurrences of $c_i$ in $t$.
\end{definition}

\begin{definition}[action $\omega\cdot (f_1,\dots,f_n)$]
Let $\bfC$ be a $\Lambda$-sorted CCC and $M$ be the $(\catccc_\Lambda)_\pp^\Op$-model corresponding to $\bfC$.
For a CCC operation $\omega : \Cl(\{T \to T'\}) \to \Cl(\{T_1 \to T_1',\dots,T_n \to T_n'\})$ and morphisms $f_i:T_i \to T_i'$ ($i=1,\dots,n$) in $\bfC$, the \emph{action} of $\omega$ on $(f_1,\dots,f_n)$ is the morphism
$$
\omega\cdot(f_1,\dots,f_n) := M\omega(f_1,\dots,f_n).
$$
\end{definition}

\begin{example}
Suppose $\Sigma = \{\square_1:T \to T', \square_2 : T' \to T''\}$
and $\Sigma' = \{c_1 : T', c_2 : T, c_3 : T \times T' \to T''\}$.
We consider $\bfC = \Cl(\Sigma')$.
For CCC operation $\omega = \llparenthesis x:T \mid_\emptyset \square_2~(\square_1~x)\rrparenthesis : \Cl(\{T \to T''\}) \to \Cl(\Sigma)$ and morphisms $f_1 = \ab(x:T \mid_\emptyset c_1)$ and $f_2 = \ab(x':T' \mid_\emptyset c_3(c_2,x'))$ in $\Cl(\Sigma')$, then $\omega\cdot (f_1,f_2)$ is equal to
\begin{align*}
&\llparenthesis x:T \mid_\emptyset \square_2~(\square_1~x)\rrparenthesis \cdot \ab(\ab(x:T \mid_\emptyset c_1), \ab(x':T' \mid_\emptyset c_3(c_2,x')))\\
&= (x:T \mid_\emptyset c_3(c_2,c_1)).
\end{align*}
\end{example}
In general, for any CCC operation $\omega = \llparenthesis x:T \mid_\emptyset t\rrparenthesis : \Cl(\{T \to T'\})\to \Cl(\{\square_1:T_1 \to T_1',\dots,\square_n : T_n \to T_n'\})$ and $f_i = (x_i:T_i \mid_\emptyset t_i) : T_i \to T_i'$ ($i=1,\dots,n$),
\begin{align*}
&\omega\cdot (f_1,\dots,f_n)=\\
&(x:T \mid_\emptyset t[(\lambda x_1:T_1.~t_1)/\square_1,\dots,(\lambda x_n:T_n.~t_n)/\square_n]).
\end{align*}

We fix a signature $\Sigma = \{c_1:T_{c_1} \to T_{c_1}',\dots,c_n : T_{c_n} \to T_{c_n}'\}$ such that each $T_{c_i}'$ is a base type.
%Consider the $\Lambda$-sorted CCC $\bfC = \Cl(\Sigma)$.

\begin{definition}
For $\Sigma' = \{\square_1 : T_1 \to T_1',\dots, \square_m : T_m \to T_m'\}$ with base types $T_1',\dots,T_m'$, a CCC operation $\omega : \Cl(T \to T') \to \Cl(\Sigma\cup \Sigma')$, and morphisms $\tau_i : T_i \to T_i'$ in $\Cl(\Sigma)$ ($i=1,\dots,m$), we define $\omega\cdot_\Sigma (\tau_1,\dots,\tau_m)$ as
\[
\omega\cdot_\Sigma(\tau_1,\dots,\tau_m) := \omega\cdot \ab(\bar c_1,\dots,\bar c_n,\tau_1,\dots,\tau_m).
\]
\end{definition}

\begin{definition}
For two CCC operations $\omega : \Cl(\{T_2 \to T_2'\}) \to \Cl(\Sigma \uplus \{T_1 \to T_1'\})$, $\omega ' : \Cl(\{T_3 \to T_3'\}) \to \Cl(\Sigma \uplus \{T_2 \to T_2'\})$, we define the CCC operation $\omega' \bullet \omega : \Cl(\{T_3 \to T_3'\}) \to \Cl(\Sigma \uplus \{T_1 \to T_1'\})$ as the composite
\begin{align*}
\Cl(T_3 \to T_3') &\xrightarrow{\omega'} \Cl(\Sigma \uplus \{T_2 \to T_2'\})\\
&\xrightarrow{\Id+ \omega} \Cl(\Sigma \uplus \{T_1 \to T_1'\}).
\end{align*}
\end{definition}
It is not difficult to show that $(\omega' \bullet \omega) \cdot_\Sigma (f_1,\dots,f_n) = \omega' \cdot_\Sigma (\omega \cdot_\Sigma (f_1,\dots,f_n))$.

\begin{lemma}\label{lem:op_ctx_sub}
Let $\omega$ be a CCC operation $\Cl(\{T \to T'\}) \to \Cl(\Sigma\cup \{\square : S \to S'\})$ with $\multiplicity{\square}{\omega} = 1$.
For any $(x:S \mid_\emptyset t : S')$ with $\beta\delta\bar\eta$-normal $t$, if $\omega\cdot_\Sigma (x:S \mid_\emptyset t) = (y:T \mid_\emptyset t' : T')$ and $t'$ is $\beta\delta\bar\eta$-normal, then $t'$ has a subterm of form $t\theta$ for some $(x:S)$-$(y:T)$ substitution $\theta$.
\end{lemma}
\begin{IEEEproof}
Suppose that $\omega = \llparenthesis y:T \mid_\emptyset s : T'\rrparenthesis$ for some $\beta\delta\bar\eta$-normal $s$.
Then, $\square$ in $s$ occurs as $\square u$ for some term $u$.
We take $\theta = \{x\mapsto u\}$ and $t'$ is obtained from $s$ by replacing $\square u$ with $t\theta$.
\end{IEEEproof}

\section{Finite derivation type}\label{sec:fdt}
In this section, we define a notion of \emph{finite derivation type (FDT)} for higher-order equation systems.
The FDT is initially defined for string rewriting systems in \cite{squier94}.

For a directed graph $G$, we write $P(G)$ for the set of paths in $G$.
For a path $p$ in $G$, $\source(p)$ denotes the source vertex of $p$ and $\target(p)$ denotes the target vertex of $p$.
The empty path (i.e., the path with length zero) on a vertex $v$ is denoted by $1_v$.
Also, for two paths $p,q$ with $\target(p) =\source(q)$, we write $p;q$ for the concatenation of $p$ and $q$.

Let $\Sigma = \{c_i : T_{c_i} \to T_{c_i}' \mid i=1,\dots,n\}$ ($T_{c_i}' \in \Lambda$, $T_{c_i}\in \Ty_\Lambda$) be a signature and $E$ be an equation system over $\Sigma$.
For $e = (\Gamma \vdash l \approx r)$, we write $e^{-1}$ for $(\Gamma \vdash r \approx l)$.
We define the set $E^{-1}$ as $\{e^{-1} \mid e \in E\}$.

In the rest of this paper, we will always assume that $E$ satisfy the following conditions:
\begin{itemize}
  \item for every $\Gamma \vdash t \approx t'$ in $E$, $\Gamma$ has length 1,
  \item $E \cap E^{-1} = \emptyset$.
%  \item $E$ does not contain any equation-in-context of form $x:T \vdash t \approx t'$ such that $t$ is $\alpha\beta\delta\eta$-equivalent to $t'$,
%  \item if $E$ contains $x:T \vdash t_1 \approx t_2$, $E$ does not contain any equation-in-context $x':T \vdash t_2' \approx t_1'$ such that $t_i$ and $t_i'[x/x']$ are $\alpha\beta\delta\eta$-equivalent for $i=1,2$.
\end{itemize}
Note that any equation system $E$ has an equivalent equation system $E'$ that satisfies the above conditions.

We define a directed graph $G_1=G_1(\Sigma,E)$ as follows.
\begin{definition}
The graph $G_1 = G_1(\Sigma,E)$ \emph{associated with} $\Sigma$ and $E$ is a directed graph such that
\begin{itemize}
  \item the set of vertices is $\Mor(\Cl(\Sigma))$,
  \item $G_1$ has an arc $(\omega,\vareqn)$ for each CCC operation
$$
\omega : \Cl(T_2 \to T_2') \to \Cl(\Sigma \uplus \{\square : T_1 \to T_1'\})
$$
with $\multiplicity{\square}{\omega}=1$ and each equation-in-context
$$\vareqn = \ab(x:T_1 \vdash l \approx r : T_1') \in E \cup E^{-1}.$$
The source and target of such $(\omega,\vareqn)$ are
\begin{align*}
\source(\omega,\vareqn) &= \omega\cdot_\Sigma (x:T_1 \mid_\emptyset l),\\
\target(\omega,\vareqn) &= \omega\cdot_\Sigma (x:T_1 \mid_\emptyset r).
\end{align*}
\end{itemize}
We call $(\omega,\vareqn)$ \emph{positive} if $\vareqn \in E$ and \emph{negative} if $\vareqn \in E^{-1}$.
\end{definition}

%Figure \ref{fig:graph_ex} shows a small part of the graph associated with $(\Sigma_1, R_1)$ given in \ref{ssec:example}.
%Note that it shows only positive arcs.

Notice that $G_1$ has an arc from $\tau_1 : T_1 \to T_1'$ to $\tau_2 : T_2 \to T_2'$ only if $T_1 = T_1'$ and $T_2 = T_2'$.
Therefore, $G_1$ has a connected component for each pair $(T,T')$ of types.
For a path $p\in P(G_1)$, we say that $p$ has \emph{type} $T \to T'$ if $p$ is in the connected component for $(T,T')$. 

%\begin{figure*}[bth]
%\begin{center}
%\includegraphics[scale=0.8]{graph_ex.pdf}
%\end{center}
%\caption{\label{fig:graph_ex} Associated graph}
%\end{figure*}

Let $\omega : \Cl(\{T_2 \to T_2'\}) \to \Cl(\Sigma \uplus \{\square : T_1 \to T_1'\})$ be a CCC operation, $a = (\omega,\vareqn)$ be an arc in $G_1$ from $\source(a) = (x:T \mid_\emptyset t : T')$ to $\target(a) = (x:T \mid_\emptyset t' : T')$ and suppose that $t,t'$ are $\beta\delta\bar\eta$-normal.
By Lemma \ref{lem:op_ctx_sub},  the existence of such arc $a$ implies that the term-in-context $x:T \vdash t$ is rewritten by to $x:T \vdash t'$ by $\vareqn = (x:T \vdash l \approx r)$ as
\(
t = t''[l\theta/x'] \to t''[r\theta/x'] = t'.
\)
In particular, $G_1$ has an edge from $(\Gamma \mid_\emptyset t)$ to $(\Gamma \mid_\emptyset t')$ if $\Gamma \vdash t$ can be rewritten to $\Gamma \vdash t'$ by an equation-in-context in $E$.

Let $P_+(G_1)$ denote the set of paths $a_1;\dots;a_k$ in $G_1$ such that $a_i$ is positive for each $i=1,\dots,k$.

We say that two vertices $\tau_1,\tau_2$ are \emph{joinable} if there exist two paths $p_1,p_2 \in P_+(G_1)$ such that $\source(p_i)=\tau_i$ ($i=1,2$) and $\target(p_1) = \target(p_2)$.

\begin{definition}
For $\omega : \Cl(\{T_2 \to T_2'\}) \to \Cl(\Sigma \uplus \{T_1 \to T_1'\})$, $\omega':\Cl(\{T_3 \to T_3'\}) \to \Cl(\Sigma \uplus \{T_1 \to T_2'\})$ and an arc $a = (\omega,e)$, we define $\omega'\cdot_\Sigma a$ as the arc $(\omega'\bullet \omega, e)$.
Also, for a path $p=a_1;\dots;a_m$, define $\omega\cdot_\Sigma p$ as the path $(\omega \cdot_\Sigma a_1);\dots;(\omega\cdot_\Sigma a_m)$.
\end{definition}

\begin{definition}[critical branching, disjoint branching]
Let $a_1 = (\omega_1,\vareqn_1)$ and $a_2 = (\omega_2,\vareqn_2)$ be two arcs such that $\source(a_1) = \source(a_2)$.
\begin{itemize}
  \item The pair $(a_1,a_2)$ is called a \emph{critical branching} if the peak $\target(a_1) \xleftarrow{\vareqn_1} \source(a_1)=\source(a_2) \xrightarrow{\vareqn_2} \target(a_2)$ is a critical peak.
  \item $a_1$ and $a_2$ \emph{overlap} if there exist CCC operation $\omega$, arcs $a_1'$, $a_2'$ such that $(a_1',a_2')$ is a critical branching, $\multiplicity{\square}{\omega}=1$ and $\omega\cdot_\Sigma a_1' = a_1$, $\omega\cdot_\Sigma a_2' = a_2$.
  \item The pair $(a_1,a_2)$ is called a \emph{disjoint branching} if $a_1$ and $a_2$ do not overlap.
\end{itemize}
\end{definition}

For a CCC operation $\omega = \llparenthesis y: T \mid_\emptyset s : T'\rrparenthesis : \Cl(\{T \to T'\}) \to \Cl(\Sigma \uplus \{\square : S \to S'\})$ such that $\multiplicity{\square}{\omega} = 1$ and $s$ is $\beta\delta\bar\eta$-normal, we define the position $\fp(\omega)$ as  the position of $\square u$ in $s$ for a term $u$.

%Let $\mathrm{Br}$ be the set of pairs $(a_1,a_1')$ of positive arcs such that $\sigma(a_1) = \sigma(a_1')$.

%For each $i=1,2$, let $\vareqn_i = (x_i:T_i \vdash l_i \approx r_i : T_i')$ be an equation-in-context in $E$, $\omega_i$ be an CCC operation with $\multiplicity{\square}{\omega} = 1$, and $a_i = (\omega_i,\vareqn_i)$ be an arc.
%Suppose that $(a_1,a_2)$ is a disjoint branching, and $\omega_i\cdot (x_i:T_i \mid_\emptyset l_i)$ is written as $\llparenthesis x_i':T_i'' \mid_\emptyset t_i : T_i''' \rrparenthesis$ for some $\beta\delta\bar\eta$-normal term $t_i$.
%By an argument similar to the proof of Critical Pair Lemma \cite{mayr1998higher}, either of the following holds:
%\begin{enumerate}
%  \item $\fp(\omega_1)$ and $\fp(\omega_2)$ are disjoint,
%  \item $\fp(\omega_2)$ is below $\fp$ for some position $\fp$ such that $t_1|_\fp$ is $y~u$ for some free variable $y$ and a term $u$,
%  \item the same with (2) but the subscript 1 and 2 are reversed.
%\end{enumerate}

The following is a formulation of Critical Pair Lemma in terms of the associated graph.
\begin{lemma}
Let $R$ be a PRS.
Let $a_1=(\omega_1,\vareqn_1)$ and $a_2 = (\omega_2,\vareqn_2)$ be positive arcs that have the same source $\source(a_1) = \source(a_2)$.
Then, either $a_1$ and $a_2$ overlap or $\target(a_1)$ and $\target(a_2)$ are joinable.
\end{lemma}
\begin{IEEEproof}
Let $\vareqn_i = (x_i : T_i \vdash l_i \approx r_i : T_i')$ for $i=1,2$.
Suppose that $(a_1,a_2)$ is a disjoint branching and show that $\target(a_1)$ and $\target(a_2)$ are joinable.

We consider the following cases.

Case (1): $\fp(\omega_1)$ and $\fp(\omega_2)$ are disjoint.
For $i = 1,2$ we write $\bar i$ for $3-i$.
There exists an edge $a_i' = (\omega_i',\vareqn_{\bar i})$ that rewrites $\omega_i\cdot (x:T_i\mid_\emptyset r_i)$ by $\vareqn_{\bar i}$ at $\fp(\omega_{\bar i})$.
Then, $a_1'$ and $a_2'$ join $\target(a_1)$ and $\target(a_2)$.

Case (2): $\fp(\omega_1)$ and $\fp(\omega_2)$ are not disjoint.
Without loss of generality, we assume $\fp(\omega_2) \succ \fp(\omega_1)$.
Since $a_1$ and $a_2$ do not overlap,
$\fp(\omega_2) \succ \fp$ for some position $\fp$ such that $t_1|_\fp$ is $$(\pr_{i_1}\dots (\pr_{i_m}x_1)\dots)~y_1~\dots~y_k$$ for some distinct bound variables $y_1,\dots,y_k$ and $i_1,\dots,i_m \in \{1,2\}$.
Here, $x_1$ is the free variable in $l_1\approx r_1$.

Suppose that $(\pr_{i_1}\dots (\pr_{i_m}x_1)\dots)$ occurs $M$ times in $l_1$ and $N$ times in $r_1$.
If $\omega_1\cdot_\Sigma (x_1 : T_1 \mid_\emptyset r_1) = (x:T \mid_\emptyset t_1)$ for a $\beta\delta\bar\eta$-normal $t_1$, then $t_1$ has the subterms $l_2\theta_2$ at $N$ disjoint positions.
Let $p_1$ be the path of length $N$ that rewrites those subterms in $\target(a_1)$ by $\vareqn_2$ in an arbitrary order.

Also, if $\omega_2\cdot_\Sigma (x_2:T_2 \mid_\emptyset r_2) = (x:T \mid_\emptyset t_2)$ for a $\beta\delta\bar\eta$-normal $t_2$, then $t_2$ has the subterms $l_2\theta_2$ at $M-1$ disjoint positions.
Let $p_2$ be the path of length $M$ that rewrites those subterms in $\target(a_2)$ by $\vareqn_2$ and then rewrites the obtained term by $\vareqn_1$.
Then, $p_1$ and $p_2$ join $\target(a_1)$ and $\target(a_2)$.
\end{IEEEproof}

\begin{corollary}
Suppose that $E$ is a PRS.
The PRS $E$ is locally confluent if and only if for any critical branching $(a,a')$, $\target(a)$ and $\target(a')$ are joinable.
\end{corollary}

Let $P^{(2)}(G)$ denote the set of pairs of paths $(p,q)$ in $G$ such that $\source(p)= \source(q)$, $\target(p) = \target(q)$.
We write $p\parallel q$ for the pair $(p,q)$ in $P^{(2)}(G)$.
We define two subsets $D$, $I$ of $P^{(2)}(G)$.

\begin{definition}[disjoint derivation]
Let $(a_1,a_2)$ be a disjoint branching.
Let $p_1,p_2$ be the paths joining $\target(a_1)$ and $\target(a_2)$ that is constructed in the proof of the previous lemma.
The pair of paths $a_1;p_1\parallel a_2;p_2$ is called a \emph{disjoint derivation}.
Let $D$ denote the set of disjoint derivations.
\end{definition}

We define a subset $I$ of $P^{(2)}(G_1)$ as
\begin{align*}
&\ab\{ \ab(a;a^{-1} \parallel 1_v) \mid a \text{ is an arc of $G_1$}, \source(a) = v\} %\\
%&\cup \ab\{ (\omega,e)\parallel 1_v\mid \multiplicity{\square}{\omega} = 0, \source((\omega,e)) = v \}.
\end{align*}

\begin{definition}
An equivalence relation ${\simeq} \subset P^{(2)}(G_1)$ is called a \emph{homotopy relation} if
\begin{itemize}
  \item $D\cup I \subset {\simeq}$,
  \item if $p \simeq q$, then $\omega\cdot_\Sigma p \simeq \omega\cdot_\Sigma q$,
  \item if $p \simeq q$, then $r_1 ; p; r_2 \simeq r_1; q; r_2$.
\end{itemize}
Here, $p \simeq q$ is pronounced as ``$p$ is homotopic to $q$''.
\end{definition}
Obviously, $P^{(2)}(G_1)$ is a homotopy relation.

\begin{definition}
For a homotopy relation $\simeq$, a subset $B \subset {\simeq}$ is called a \emph{homotopy basis} of $\simeq$ if $\simeq$ is the smallest homotopy relation that contains $B$.
In that case we write $\simeq_B$ for $\simeq$.
Also, we say that $(\Sigma,E)$ has \emph{finite derivation type (FDT)} if there exists a finite homotopy basis of $P^{(2)}(G_1)$.
\end{definition}

In fact, FDT is a property of CCCs rather than equational theories, i.e., under the condition $\Cl(\Sigma,E)\cong \Cl(\Sigma',E')$, $(\Sigma,E)$ has FDT if and only if so does $(\Sigma',E')$.
The proof is done in a similar way to the case for string rewriting systems, but
we do not use this fact in this paper so omit the details.

\begin{definition}
For $B \subset P^{(2)}(G_1)$, we define a relation ${\Rightarrow_B} \subset P^{(2)}(G_1)$ as follows:
$p \Rightarrow_B q$ if there exist $(p'\parallel q') \in B \cup D \cup I$, a CCC operation $\omega$, paths $r_1,r_2 \in P(G_1)$ such that
\begin{enumerate}
  \item $\target(r_1) = \source(\omega\cdot_\Sigma p')$, $\source(r_2) = \target(\omega\cdot_\Sigma p')$,
  \item $p=r_1; (\omega\cdot_\Sigma p'); r_2$, $q = r_1; (\omega\cdot_\Sigma q'); r_2$.
\end{enumerate}
\end{definition}
It is easy to show that the reflexive, transitive, symmetric closure of $\Rightarrow_B$ coincides with $\simeq_B$.

For $B\subset P^{(2)}(G_1)$,
let $B^{-1} = \{(q\parallel p) \mid (p\parallel q) \in B\}$.
We assume $B \cap B^{-1} = \emptyset$.
We define a directed graph $G_2(B)$ as follows:
\begin{itemize}
  \item the set of vertices of $G_2(B)$ is $P(G_1)$,
  \item $G_2(B)$ has an arc $\alpha =(\omega,r_1,(p\parallel q),r_2)$ for each parallel paths $p\parallel q \in B\cup B^{-1}$ of type $T_1 \to T_1'$,
CCC operation
\[
\omega : \Cl(\{T_2 \to T_2'\}) \to \Cl(\Sigma\uplus \{\square : T_1 \to T_1'\})
\]
with $\multiplicity{\square}{\omega} = 1$,
and paths $r_1,r_2$ with $\target(r_1) = \source(\omega\cdot_\Sigma p)$, $\source(r_2) = \target(\omega\cdot_\Sigma p)$ and $r_1;\omega\cdot_\Sigma p; r_2$ has type $T_2 \to T_2'$.
  \item The source and target of $\alpha$ is given as $\source(\alpha) = r_1;(\omega\cdot_\Sigma p);r_2$, $\target(\alpha) = r_1;(\omega\cdot_\Sigma q);r_2$.
\end{itemize}

From now, we assume $E = R$ is a complete PRS that is finite as a set.
For $\tau = (x:T \mid_\emptyset t)$, we write $\hat{\tau}$ for $\ab(x:T \mid_\emptyset \hat t)$ where $\hat t$ is the $R$-normal form of $t$.
For each vertex $\tau$, we choose a path $p(\tau)$ in $G_1$ from $\tau$ to $\hat \tau$.

Let $C=C(\Sigma,R) \subset P^{(2)}(G_1)$ be the set consisting of
\[
\ab(a_1;p(\target(a_1))\parallel a_2;p(\target(a_2)))
\]
for each critical branching $(a_1,a_2)$.

\begin{lemma}
Let $a_1,a_2$ be positive arcs with the same source.
Then, there exist $q_1,q_2 \in P_+(G_1)$ such that $\target(a_i) = \source(q_i)$ ($i=1,2$), $\target(q_1) = \target(q_2)$, and $a_1;q_1 \simeq_C a_2;q_2$.
\end{lemma}
\begin{IEEEproof}
Let $a_i = (\omega_i,e_i)$ ($i=1,2$).
If $(a_1,a_2)$ is a disjoint branching, then we can take $q_1,q_2$ such that $(a_1;q_1, a_2;q_2)$ is a disjoint derivation.
If $a_i = \omega\cdot_\Sigma a_i'$ and $(a_1',a_2')$ is a critical branching for some $\omega$, $a_i'$ ($i=1,2$), by the definition of $C$, there exist paths $q_1',q_2'$ such that $a_1;\omega\cdot_\Sigma q_1' \simeq_C a_2;\omega\cdot_\Sigma q_2'$.
\end{IEEEproof}

\begin{lemma}
Let $p_1,p_2\in P_+(G_1)$ be paths such that $\source(p_1)=\source(p_2) = \tau$, $\target(p_1)=\target(p_2) = \hat \tau$.
Then, $p_1 \simeq_C p_2$.
\end{lemma}
\begin{IEEEproof}
Let $\tau = (x:T \mid_\emptyset t)$.
We proceed by well-founded induction on $t$ with respect to $\to_R$.
If $t$ is normal, then $\tau=\hat \tau$ and so $p_1, p_2$ are homotopic to $1_\tau$.

Suppose that $t$ is not normal.
Then we have arcs $a_1,a_2$ and paths $p_1',p_2'$ such that $p_1 = a_1;p_1'$ and $p_2 = a_2;p_2'$.
By the previous lemma, there exist $q_1,q_2 \in P_+(G_1)$ such that $a_1;q_1 \simeq_C a_2;q_2$.
By the induction hypothesis, we have $p_1' \simeq_C q_1$, $p_2' \simeq_C q_2$.
Thus $p_1 = a_1;p_1' \simeq_C a_1;q_1 \simeq_C a_2;q_2 \simeq_C a_2;p_2' = p_2$.
\end{IEEEproof}

\begin{lemma}
Let $p \in P(G_1)$ be a path from $\tau_1$ to $\tau_2$ and $p_1,p_2 \in P_+(G_1)$ be paths from $\tau_1,\tau_2$ to $\hat \tau_1,\hat \tau_2$, respectively.
Then, $p \simeq_C p_1; p_2^{-1}$.
\end{lemma}
\begin{IEEEproof}
We proceed by induction on the length $n$ of $p$.
If $n=0$, then $\tau_1 = \tau_2$ and $p = 1_{\tau_1} = 1_{\tau_2}$.
Since $p_1 \simeq_C p_2$ by the previous lemma, $1_{u_1} \simeq_C p_2;p_2^{-1}\simeq_C p_1; p_2^{-1}$.

If $n > 0$, there exists an arc $a$ and a path $p'$ such that $p = p';a$.
Let $\tau = \target(p')$ and $q$ be a positive path from $\tau$ to $\hat \tau$.
Then, by the induction hypothesis, we have $p' \simeq_C p_1;q^{-1}$.

If $a$ is a positive arc, then we can apply the previous lemma for $a;p_2$ and $q$, hence $a;p_2 \simeq_C q$.
If $e$ is a negative arc, then we can apply the previous lemma for $p_2$ and $a^{-1};q$, hence $p_2\simeq_C a^{-1};q$.
In either case, we conclude $p \simeq_C p_1; p_2^{-1}$.
\end{IEEEproof}

\begin{theorem}
$C=C(\Sigma,R)$ is a basis of $P^{(2)}(G_1)$.
\end{theorem}
\begin{IEEEproof}
Let $(p\parallel q)\in P^{(2)}(G_1)$, $r_1$ be a path from $\source(p)$ to $\widehat{\source(p)}$, and $r_2$ be a path from $\target(p)$ to $\widehat{\target(p)}$.
Then, by the previous lemma, we have $p \simeq_C r_1;r_2^{-1} \simeq_C q$.
\end{IEEEproof}

\begin{corollary}
$R$ has FDT if it is finite (as a set) and complete.
\end{corollary}

\section{Ringoids and modules}\label{sec:ringoid}
To define homology of CCCs, we use the notion of ringoids and modules over them.
For more detailed information of ringoids and modules, see \cite{mitchell1972}.

\begin{definition}
A \emph{ringoid} is a small $\catab$-enriched category.
In other words, a ringoid is a small category $\cR$ such that each hom-set $\Hom_\cR(X,Y)$ has an abelian group structure $(\Hom_\cR(X,Y),+,0)$ and the composition is bilinear, i.e., for any $r_1,r_2 : X \to Y$, $r_3 : Y \to Z$,
\[
r_3 \circ (r_1 + r_2) = r_3 \circ r_1 + r_3 \circ r_2,
\]
and for any $r_1 : x \to y$, $r_2,r_3 : y \to z$,
\[
(r_2+r_3 ) \circ r_1 = r_2 \circ r_1 + r_3 \circ r_1.
\]
\end{definition}

\begin{definition}
Let $\cR$ be a ringoid.
A \emph{left $\cR$-module} is a functor $M : \cR \to \catab$ such that $M(r+s) = Mr + Ms$, $M0 = 0$.
A \emph{right $\cR$-module} is a left $\cR^\Op$-module.
We often just say an \emph{$\cR$-module} for a left $\cR$-module.
\end{definition}

For a left $\cR$-module $M$, morphism $r : x \to y$ in $\cR$, and $m \in M(X)$, we write $f\cdot m$ for $Mr(m)\in M(Y)$ and call it the \emph{scalar multiplication} of $m$ by $r$.
For a right $\cR$-module $N$, morphism $r : Y \to X$ in $\cR$, and $m \in M(X)$, we write $m\cdot r$ for $Mr(m) \in M(Y)$.

\begin{definition}
For two $\cR$-modules $M,N$, an $\cR$-\emph{linear map} from $M$ to $N$ is a natural transformation from $M$ to $N$.
\end{definition}
We write $\catmodules{\cR}$ for the category of $\cR$-modules and $\cR$-linear maps.
It is well-known that $\catmodules{\cR}$ is an abelian category with enough projectives.

\begin{definition}
For an $\cR$-linear map $f : M \to N$,
\begin{itemize}
  \item the \emph{kernel} of $f$ is the $\cR$-module $\ker f$ defined as $(\ker f)(X) = \ker f_X$ for each $X \in \Ob(\cR)$ and $(\ker f)(r)(m) = Mr(m) \in \ker f_Y$ for $r : X \to Y$, $m \in MX$,
  \item the \emph{image} of $f$ is the $\cR$-module $\im f$ defined as $(\im f)(X) = \im f_X$ for each $X \in \Ob(\cR)$ and $(\im f)(r)(m) = Nr(m) \in \im f_Y$ for $r : X \to Y$, $m \in NX$.
\end{itemize}
\end{definition}
\begin{definition}
A (finite or infinite) sequence of $\cR$-linear maps $\dots \to M_{i+1} \xrightarrow{f_i} M_i \xrightarrow {f_{i-1}} \dots$ is \emph{exact} if $\ker f_{i-1} = \im f_i$ for each $i$.
\end{definition}

\begin{definition}
For a family $\cS$ of sets $\cS_X$ ($X \in \Ob(\cR)$),
the \emph{free $\cR$-module} generated by $\cS$, denoted $\cR\gen{\cS}$, is defined as follows:
$\cR\gen{\cS}(X)$ is the abelian group of formal sums
\[
\sum_{Y\in \Ob(\cR),a\in \cS_Y} r_a\gen{a}\quad (r_a : Y\to X)
\]
where only finite number of $r_a$ is non-zero, and the scalar multiplication is given as
\[
r\cdot\ab( \sum_{Y\in \Ob(\cR),a\in \cS_Y} r_a\gen{a}) = \sum_{Y\in \Ob(\cR),a\in \cS_Y} (r\circ r_a)\gen{a}.
\]
the free $\cR$-module can be described as the functor $\cS \mapsto \cR\gen{\cS}$ that is a left adjoint of the forgetful functor from the category of $\cR$-modules to the category $\catset^{\Ob(\cR)}$.

Also, for a family $\cE$ of subsets $\cE_{X}\subset \cR\gen{\cS}(X)$ and a $\cR$-module $M$, we say that $M$ is \emph{presented by $(\cS,\cE)$} if $M$ is isomorphic to the quotient of $\cR\gen{\cS}$ by the submodule generated by $\cE$, i.e., the smallest submodule $N$ such that $\cE_X\subset N(X)$.
Elements of $\cE_X$ are called \emph{relations}.
\end{definition}

\begin{definition}
Let $M$ be a left $\cR$-module and $N$ be a right $\cR$-module.
The \emph{tensor product} of $N$ and $M$, written $N\otimes_\cR M$, is the enriched coend $\int^X MX\otimes NX$, or explicitly, it
is the abelian group defined as the quotient of $\bigoplus_X N(X)\otimes M(X)$ by relations $(n\cdot r)\otimes m - n \otimes (r\cdot m)$ for all $r : X \to Y$, $n \in N(Y)$, $m \in M(X)$.
\end{definition}

For two left $\cR$-modules $M_1,M_2$, a right $\cR$-module $N$, any $\cR$-linear map $f:M_1 \to M_2$ extends to an abelian group homomorphism $N\otimes_\cR f : N \otimes_\cR M_1 \to N\otimes_\cR M_2$.
Moreover, $N\otimes_\cR - : \catmodules{\cR} \to \catab$ forms an additive functor.

\begin{definition}
Let $M$ be an $\cR$-module.
A \emph{free resolution} of $M$ is an exact sequence
\begin{equation}\label{eqn:freeres}
\dots \to F_2 \xrightarrow{f_2} F_1 \xrightarrow{f_1} F_0 \to M \to 0
\end{equation}
where $F_i$ is a free $\cR$-module for each $i=0,1,\dots$.
\end{definition}

From a general theorem in homological algebra (see \cite[Chapter 2]{Weibel1994} for example), given a free resolution (\ref{eqn:freeres}) of $M$,
the sequence
\[
\dots \to N\otimes_\cR F_2 \xrightarrow{N\otimes_\cR f_2} N\otimes_\cR F_1 \xrightarrow{N\otimes_\cR f_1} N\otimes_\cR F_0
\]
in $\catab$ satisfies $\ker(N\otimes_\cR f_{i}) \supseteq \im(N\otimes_\cR f_{i+1})$ for each $i$, and the quotient abelian group
\(
\ker(N\otimes_\cR f_{i})/\im(N\otimes_\cR f_{i+1})
\)
depends only on the $\cR$-module $M$, not on the choice of the free resolution.
This abelian group is written as $\Tor_i^\cR(N,M)$.

%\begin{definition}
%Let $O$ be a set.
%For a family $\cG$ of sets $\cG_{X,Y}$ indexed by $X,Y\in O$, the \emph{ringoid freely generated by} $\cG$ \emph{with coefficients in} $\bbQ$ is the ringoid $\cF_\cG$ whose set of objects is $O$ and hom-set $\Hom_{\cF_\cG}(X,Y)$ consists of formal linear combinations
%\[
%\sum_{X,Y\in O,~r_{X,Y}\in \cG_{X,Y}} q_{X,Y}r_{X,Y} \quad (q_{X,Y}\in \bbQ)
%\]
%where only finitely many $q_{X,Y}$s are non-zero.
%The ringoid structure is naturally given.
%
%Also, given a family $\cE$ of subsets $\cE_{X,Y}\subset\Hom_{\cF_\cG}(X,Y)$, the \emph{ringoid presented by $\cG$, $\cE$} is 
%\end{definition}

\section{Proof of Main Theorem}\label{sec:proof}

Let $\bfC$ be a $\Lambda$-sorted CCC.
We define a ringoid $\cU_\bfC$ and a module $\Omega_\bfC$ over it, and then we define the homology of $\bfC$ from a resolution of $\Omega_\bfC$.
The definitions of $\cU_\bfC$ and $\Omega^1_\bfC$ are obtained from a general discussion in \cite{jp06}.
\begin{definition}
We define a ringoid $\cU_\bfC$ as follows.
The set of objects is $\Ob(\cU_\bfC) = \Mor(\bfC)$, and we give the morphisms of $\cU_\bfC$ by generators and relations:
For each CCC operation $\omega : \Cl(\{T \to T'\}) \to \Cl(\{\square_1:T_1 \to T_1' ,\dots, \square_n:T_n \to T_n'\})$ and
$f_1:T_1\to T'_1,\dots,f_n:T_n \to T'_n$ in $\Mor(\bfC)$,
we have $n$ generators written as
\[
\partial_{\square_i}(\omega)_{(f_1,\dots,f_n)} \colon f_i \to \omega\cdot (f_1,\dots,f_n) \quad (i = 1,\dots,n)
\]
with coefficients in $\bbQ$,
and for each $\omega : \Cl(\{T \to T\}) \to \Cl(\{\square_1:T_1 \to T_1',\dots,\square_n : T_n \to T_n'\})$,
$\omega_i : \Cl(\{T_i \to T_i'\}) \to \Cl(\{\square'_1 : U_{1} \to U_{1}',\dots,\square_m' : U_{m} \to U_{m}' \})$ ($i = 1,\dots,n$),
and $f_{j} \colon U_{j} \to U_{j}'$ ($j=1,\dots,m$) in $\bfC$, we impose relations
\begin{align*}
&\partial_{\square_j'}([ \omega_1,\dots,\omega_n] \circ \omega)_{(f_1,\dots,f_m)} =\\
&\sum_{i=1}^n \partial_{\square_i}(\omega)_{(\omega_1\cdot (f_1,\dots,f_m),\dots, \omega_n\cdot (f_1,\dots,f_m))}\circ \partial_{\square_j'}(\omega_i)_{(f_1,\dots,f_m)}
\end{align*}
and also
\[
\partial_{\square_j}(\iota_i)_{(f_1,\dots,f_n)} =
\begin{cases}
  \Id_{f_i} & \text{if $j=i$}\\
  0 & \text{otherwise}
\end{cases}
\]
where $\iota_i : \Cl(\{\square_i : T_i \to T_i'\}) \to \Cl(\{\square_1 : T_1 \to T_1',\dots,\square_n : T_n \to T_n'\})$ is the canonical inclusion.
We call $\cU_\bfC$ the \emph{enveloping ringoid} of $\bfC$.
\end{definition}

That is, morphisms of $\cU_\bfC$ can be written as a formal sum of formal expressions
\[
q~ \partial_{\square_{i_1}}(\omega_1)_{(f_{1,1},\dots,f_{n_1,1})} \circ \dots \circ \partial_{\square_{i_m}}(\omega_m)_{(f_{1,m},\dots,f_{n_m,m})}
\]
where $q \in \bbQ$ is a rational number, $\omega_j$s are CCC operations, $f_{i,j}$s are morphisms of $\bfC$, and $\circ$ is the formal composition.
Such formal sums satisfy the imposed relations.
Here, we can compose two generators as $\partial_{\square_j}(\omega')_{(g_1,\dots,g_m)} \circ \partial_{\square_i}(\omega)_{(f_1,\dots,f_n)}$ if and only if $\omega\cdot (f_1,\dots,f_n) = g_j$.

For any function symbol $c : T \to T'$ in a signature $\Sigma$, recall that $\bar c$ is the shorthand for $(x:T \mid_\emptyset cx)$ in $\Cl(\Sigma)$.

We can intuitively think of $\partial_X(\omega)_{(f_1,\dots,f_n)}$ as the partial derivative of $\omega$ along $X$ at $(f_1,\dots,f_n)$ in the following way:
\begin{example}
Suppose that CCC operation $\omega : \Cl(\{T \to T\}) \to \Cl(X : T \to T)$ is written as $\llparenthesis X \circ X \rrparenthesis$.
Then, since $\llparenthesis X \circ X \rrparenthesis = [\llparenthesis \bar Y_1 \rrparenthesis, \llparenthesis \bar Y_2\rrparenthesis] \circ \llparenthesis \bar Y_1 \circ \bar Y_2 \rrparenthesis$, we have
\begin{align*}
\partial_{X}(\llparenthesis \bar X \circ \bar X \rrparenthesis)_{(f)}
&= \partial_{Y_1}(\llparenthesis \bar Y_1 \circ \bar Y_2 \rrparenthesis)_{(f,f)} \circ \partial_X(\llparenthesis \bar X \rrparenthesis)_{(f)}\\
&\quad + \partial_{Y_2}(\llparenthesis \bar Y_1 \circ \bar Y_2 \rrparenthesis)_{(f,f)} \circ \partial_X(\llparenthesis \bar X \rrparenthesis)_{(f)}\\
&= \partial_{Y_1}(\llparenthesis \bar Y_1 \circ \bar Y_2 \rrparenthesis)_{(f,f)} + \partial_{Y_2}(\llparenthesis \bar Y_1 \circ \bar Y_2 \rrparenthesis)_{(f,f)}.
\end{align*}
\end{example}

The following is not difficult to show.
\begin{lemma}\label{lem:deriv_perm}
For any CCC operation $\omega: \Cl(\{T \to T'\})\to \Cl(\{\square_1 : T_1\to T_1',\dots,\square_m : T_m \to T_m'\})$,
morphisms $f_1 : T_1 \to T_1',\dots,f_m : T_m \to T_m'$ in $\bfC$, a permutation $\sigma$ on $\ab\{1,\dots,m\}$, if $\sigma(j) = i$ for some $i,j\in \ab\{1,\dots,m\}$, then
\[
\partial_{\square_i}\ab([\iota_{\sigma(1)},\dots,\iota_{\sigma(m)}]\circ\omega)_{(f_1,\dots,f_n)} = \partial_{\square_j}(\omega)_{\ab(f_{\sigma(1)},\dots,f_{\sigma(m)})}.
\]
\end{lemma}

\begin{definition}
We define a $\cU_\bfC$-module $\Omega_\bfC^1$ as follows.
For each $f : T \to T'$ in $\bfC$, it has a generator $df$ in $\Omega_\bfC^1(f)$,
for each $f_i : T_i \to T'_i$ ($i=1,\dots,m$) and $\omega : \Cl(\{T \to T'\}) \to \Cl(\{\square_1 : T_1 \to T_1',\dots,\square_n : T_m \to T_m'\})$, it has a relation
\[
d(\omega\cdot (f_1,\dots,f_m)) = \sum_{i=1}^m \partial_{\square_i}(\omega)_{(f_1,\dots,f_m)}df_i.
\]
\end{definition}
Those who may be familiar with the notion of differential forms may notice that the elements of $\Omega_\bfC^1(f)$ can be intuitively thought of as differential 1-forms.

Suppose that CCC $\bfC$ is presented by $(\Sigma,E)$ for a $\Lambda$-sorted signature $\Sigma=\ab\{c_1 : T_{c_1} \to T_{c_1}',\dots,c_n:T_{c_n} \to T_{c_n}' \}$ where each $T_i'$ is a base type.
Also, as we did in Section \ref{sec:mainthm}, we assume that for each $\Gamma \vdash l \approx r$ in $E$, the multisets of free variables in $l$ and $r$ are the same.
We consider $\Sigma$ as the functor $\Mor(\bfC) \to \catset$ that maps $f \in \Mor(\bfC)$ to $\Sigma_f = \ab\{c_i \in \Sigma \mid \ab(x:T_{c_i} \mid_E c_i~x) = f\}$.
Also, consider $E$ as the functor $\Mor(\bfC) \to \catset$ that maps $f \in \Mor(\bfC)$ to $$E_f = \ab\{ \Gamma \vdash l \approx r \in E \mid \ab(\Gamma \mid_E l) = f\}.$$

We write $\tilde c_i$ for $(x:T_{c_i} \mid_E c_i~x)$.

\begin{lemma}\label{lem:deriv_canon}
For any CCC operation $\omega : \Cl(\{T \to T'\}) \to \Cl(\{\square_1 : T_1\to T_1',\dots,\square_m : T_m\to T_m'\})$, morphisms $f_1:T_1\to T_1',\dots,f_m:T_m \to T_m'$ in $\bfC$, and index $i\in \{1,\dots,m\}$,
there exists a CCC operation $\omega' : \Cl(\{T \to T'\}) \to \Cl(\Sigma \uplus \{\square : T_i \to T_i'\})$ such that
\[
\partial_{\square_i}(\omega)_{(f_1,\dots,f_m)} = \partial_{\square}(\omega')_{(\tilde c_1,\dots,\tilde c_n,f_i)}.
\]
\end{lemma}
\begin{IEEEproof}
By Lemma \ref{lem:deriv_perm}, we can assume $i=m$.
Suppose $f_j = (x:T_j \mid_E t_j : T_j')$ for $j=1,\dots,n$.
By taking $\omega' = [[\iota_1,\dots,\iota_n]\circ \llparenthesis f_1\rrparenthesis, \dots, [\iota_1,\dots,\iota_n]\circ \llparenthesis f_n \rrparenthesis, \iota_{n+1}] \circ \omega$,
we have
\begin{align*}
&\partial_{n+1}(\omega')_{\ab(\tilde c_1,\dots,\tilde c_n,f_m)}\\
&= \sum_{i=1}^{m-1} \partial_i(\omega)_{(f_1,\dots,f_m)}\circ \partial_{n+1}([\iota_1,\dots,\iota_n]\circ\llparenthesis f_i \rrparenthesis)_{\ab(\tilde c_1,\dots,\tilde c_n,f_m)}\\
&\quad + \partial_{m}(\omega)_{(f_1,\dots,f_m)}\circ \partial_{n+1}( \iota_{n+1})_{\ab(\tilde c_1,\dots,\tilde c_n,f_m)}\\
&= \partial_{m}(\omega)_{(f_1,\dots,f_m)}.
\end{align*}
\end{IEEEproof}

We write $\partial(\omega)_{(f)}$ or just $\partial(\omega)$ for $\partial_{\square}(\omega)_{\ab(\tilde c_1,\dots,\tilde c_n,f)}$ for $\omega : \Cl(\{T \to T'\}) \to \Cl(\Sigma \uplus \{\square : U \to U'\})$.

The following lemma can be proved by a simple computation.
\begin{lemma}\label{lem:deriv_comp}
\(
\partial(\omega)_{(\omega'\cdot_\Sigma f)} \circ \partial(\omega')_{(f')} =
\partial(\omega\bullet \omega')_{(f')}
\)
\end{lemma}

Also, by Lemma \ref{lem:deriv_canon} and Lemma \ref{lem:deriv_comp}, we have
\begin{lemma}
Any morphism $f \to f'$ in $\cU_\bfC$ can be written as
$
\partial(\omega_1)(f) + \dots + \partial(\omega_m)(f)
$
for some $\omega_i$ with $\omega_i\cdot_\Sigma f = f'$ ($i=1,\dots,m$).
\end{lemma}

%We can check by a simple computation that any (either left or right) $\cU_\bfC$-module $M$ satisfies $M_{\ab<f_1,f_2>} \cong M_{f_1}\times M_{f_2}$ and $M_{\lambda f} \cong M_f$,
%and then, to define a $\cU_\bfC$-module $M$, it suffices to define $M_f$ for each $f : X \to Y$ with $Y \in \Lambda$ and scalar multiplications $M_f \to M_{f'}$ for $f : X \to Y$, $f' : X' \to Y'$ ($Y,Y' \in \Lambda$).
\begin{definition}\label{def:coefmulti}
For a morphism $f = (x:T_1\times \dots \times T_n \mid_E t)$ in $\bfC$, we define a natural number $N_i(f)$ as the number of occurrences of $x_i$ in the normal form of $t[(x_1,\dots,x_n)/x]$.
Note that this number is well-defined because of our assumption on $E$.
\end{definition}

\begin{definition}\label{def:coef}
We define the right $\cU_\bfC$-module $\cQ$ as follows.
Let $\cQ_f = \bbQ$ for each $f : X \to Y$ ($Y \in \Lambda$) in $\bfC$,
$\cQ_{\ab<f_1,f_2>} = \cQ_{f_1}\times \cQ_{f_2}$ for each $f_i : X \to X_i$ in $\bfC$ ($i=1,2$),
$\cQ_{\lambda f} = \cQ_f$ for each $f : X \times Y \to Z$ in $\bfC$,
and define the scalar multiplication as
\begin{align*}
&r\cdot \partial_{\square_1}\llparenthesis \square_1 \circ \square_2 \rrparenthesis_{(f,g)} =  r,\\
&r\cdot \partial_{\square_2}\llparenthesis \square_1 \circ \square_2 \rrparenthesis_{(f,g)} =  (rN_{1}{(f)},\dots,rN_{n}(f))\\
&\quad (f : T_1\times \dots \times T_n \to T,~ T \in \Lambda),\\
&(r_1,r_2)\cdot \partial_{\square_i}\llparenthesis \ab<\square_1, \square_2> \rrparenthesis_{(f_1,f_2)} = r_i \quad (i=1,2),\\
&r \cdot \partial_{\square}\llparenthesis \lambda\square \rrparenthesis_{(f)} = r
\end{align*}
The scalar multiplication of $\partial_{\square_i}(\omega)_{(f_1,\dots,f_n)}$ for general $\omega$ is determined from the above and the relations of $\cU_\bfC$.
%k$r\cdot \partial(\omega)_{(f)} = \cQ(\partial(\omega))(r) = r\multiplicity{n+1}{\omega}$ ($r \in \bbQ$).
%kNote that the well-definedness of this scalar multiplication follows from the next lemma.
\end{definition}

%\begin{lemma}
%For any CCC operations $\omega : \Cl(\{T \to T'\}) \to \Cl(\square_1 : T_1 \to T_1',\dots,\square_m : T_m \to T_m')$, $\omega_i : \Cl(\{T_i\to T_i'\}) \to \Cl(\{\square_{1}' : U_{1} \to U_{1}',\dots, \square_k' : U_k \to U_{k}'\})$ ($i=1,\dots,m$), we have
%\[
%\multiplicity{\square_j'}{\ab[ \omega_1,\dots,\omega_m]\circ \omega} = \sum_{i=1}^m \multiplicity{\square_i}{\omega}\multiplicity{\square_j'}{\omega_i}
%\]
%\end{lemma}
%\begin{IEEEproof}
%If $\omega$ is determined by a term $t$ and $\omega_i$ is determined by a term $t_i$, then $[\omega_1,\dots,\omega_m]\circ \omega$ is determined by the term $t[t_1/\square_1,\dots,t_m/\square_m]$.
%So, the formula follows from the definition of $\multiplicity{\square}{\omega}$.
%\end{IEEEproof}

%\begin{lemma}
%For any family $\cS$ of sets $\cS_f$ indexed by $f \in \Mor(\bfC)$,
%$\cQ\otimes_{\cU_\bfC} \cU_\bfC S$ is the $\bbQ$-vector space generated by the elements of $\biguplus_{f\in\Mor(\bfC)} \cS_f$.
%\end{lemma}

We define the homology of $\bfC$ (with coefficients in $\cQ$) as $H_n(\bfC) = \Tor_n^{\cU_\bfC}(\cQ,\Omega_\bfC^1)$.

Let $B$ be a homotopy basis of $P^{(2)}(\Sigma,E)$.
From now, we construct a partial free resolution of $\Omega_\bfC^1$
\begin{equation}
\label{eqn:resolution_omega}
\cU_\bfC B \xrightarrow{\delta_2} \cU_\bfC E \xrightarrow{\delta_1} \cU_\bfC \Sigma \xrightarrow{\epsilon} \Omega_\bfC^1 \to 0.
\end{equation}
The idea of construction is based on \cite{cremanns1994} for string rewriting systems.

To define $\delta_i$ ($i=1,2$) and show the exactness, we construct auxiliary modules and maps.

Let $\cP_0$, $\cP_1$, $\cP_2$ be the family of sets indexed by $\Mor(\bfC)$, i.e., functors from the discrete category $\Mor(\bfC)$ to $\catset$, defined as follows:
\begin{align*}
\cP_0(f) &= \ab\{\tau \in \Mor (\Cl(\Sigma))\mid [\tau]_E = f\},\\
\cP_1(f) &= \ab\{p \in P(G_1(\Sigma,E)) \mid [\source(p)]_E = f \},\\
\cP_2(f) &= \ab\{ \varpi \in P(G_2(B)) \mid  [\source(\source(\varpi))]_E = f \}
\end{align*}
where, for a morphism $\tau$ in $\Cl(\Sigma)$, $[\tau]_E$ is the equivalence class of $\tau$ in $\Cl(\Sigma,E)\cong \bfC$.

For any $\vareqn = (\Gamma \vdash l \approx r)$ in $E_f$, let $\psi_{1,\mathrm{L}}(\vareqn) = (\Gamma \mid l)$ and $\psi_{1,\mathrm{R}}(\vareqn) = (\Gamma \mid r)$.
Then, $\psi_{1,\mathrm L}$ and $\psi_{1,\mathrm R}$ extend to $\cU_\bfC$-linear maps
\[
\psi_{1,\mathrm L} : \cU_\bfC E \to \cU_\bfC \cP_0, \quad \psi_{1,\mathrm R} : \cU_\bfC E \to \cU_\bfC \cP_0.
\]

Also, for any $p \parallel q$ in $B_f$, let $\psi_{2,\mathrm{L}}(p \parallel q) = p$ and $\psi_{2,\mathrm{R}}(p \parallel q) = q$.
Then, $\psi_{2,\mathrm L}$ and $\psi_{2,\mathrm R}$ extend to $\cU_\bfC$-linear maps
\[
\psi_{2,\mathrm L} : \cU_\bfC B \to \cU_\bfC \cP_1, \quad \psi_{2,\mathrm R} : \cU_\bfC B \to \cU_\bfC \cP_1.
\]

We are going to define two $\cU_\bfC$-linear maps
\[
\varphi_0 : \cU_\bfC\cP_0 \to \cU_\bfC \Sigma, \quad \varphi_1 : \cU_\bfC\cP_1 \to \cU_\bfC E
\]
and then define $\delta_1$ and $\delta_2$ as
\[
\delta_1 = \varphi_1 \circ (\psi_{1,\mathrm L} - \psi_{1,\mathrm R}), \quad \delta_2 = \varphi_2 \circ (\psi_{2,\mathrm L} - \psi_{2,\mathrm R}).
\]
The following diagram shows the maps we have defined so far or are going to define:
\[
\begin{tikzcd}
\cU_\bfC\cP_2 \arrow[d,"\varphi_2"] & \cU_\bfC\cP_1 \arrow[d,"\varphi_1"] & \cU_\bfC\cP_0 \arrow[d,"\varphi_0"] &\\
\cU_\bfC B \arrow[r,  "\delta_2"] \arrow[ru, shift left, "\psi_{2,\mathrm L}"] \arrow[ru, shift right, "\psi_{2,\mathrm R}"', near end] & \cU_\bfC E \arrow[r, "\delta_1"]  \arrow[ru, shift left, "\psi_{1,\mathrm L}"] \arrow[ru, shift right, "\psi_{1,\mathrm R}"', near end] & \cU_\bfC \Sigma \ar[r, "\epsilon"]  & \Omega_\bfC^1
\end{tikzcd}
\]

First, we define $\cU_\bfC$-linear maps $\epsilon : \cU_\bfC \Sigma \to \Omega_\bfC^1$ and $\varphi_0 : \cU_\bfC \cP_0 \to \cU_\bfC \Sigma $ by
\[
\epsilon\ab(\gen{c_i}) = d\tilde c_i, \quad
\varphi_0(\tau) = \sum_{i=1}^n \partial_i\llparenthesis \tau \rrparenthesis_{(\tilde c_1,\dots,\tilde c_n)\gen{c_i}}.
\]
Then, we define $\delta_1$ by $$\delta_1(\gen{\Gamma \vdash l \approx r}) = \varphi_0(\Gamma \mid_\emptyset l) - \varphi_0(\Gamma \mid_\emptyset r).$$

For any morphism $f = (x:T \mid_E t)$ in $\bfC \cong \Cl(\Sigma,E)$, 
since $f= \llparenthesis x:T \mid_\emptyset t \rrparenthesis \cdot \ab(\tilde c_1,\dots,\tilde c_n)$, we have
\begin{align*}
df &= d(\llparenthesis x:T \mid_\emptyset t \rrparenthesis \cdot \ab(\tilde c_1,\dots,\tilde c_n))\\
&= \sum_{i=1}^n \partial_{\square_i}\llparenthesis x:T \mid_\emptyset t \rrparenthesis_{(\tilde c_1,\dots,\tilde c_n)}d\tilde c_i.
\end{align*}
Therefore, $\Omega^1_\bfC$ is generated by $d\tilde c_1,\dots,d\tilde c_n$.

With these generators, $\Omega^1_\bfC$ has a relation
\begin{align*}
&\sum_{i=1}^n \partial_{\square_i}\llparenthesis x:T \mid_\emptyset t_1 \rrparenthesis_{\ab(\tilde c_1,\dots,\tilde c_n)}d\tilde c_i\\
 &= \sum_{i=1}^n \partial_{\square_i}\llparenthesis x:T \mid_\emptyset t_2 \rrparenthesis_{\ab(\tilde c_1,\dots,\tilde c_n)}d\tilde c_i
\end{align*}
for each pair of terms $t_1,t_2$ such that $x:T \vdash t_1\approx_E t_2$.
Moreover, the relation above is derivable from relations
\begin{align*}
&\sum_{i=1}^n \partial_{\square_i}\llparenthesis x:T \mid_\emptyset l \rrparenthesis_{\ab(\tilde c_1,\dots,\tilde c_n)}d\tilde c_i\\
& = \sum_{i=1}^n \partial_{\square_i}\llparenthesis x:T \mid_\emptyset r \rrparenthesis_{\ab(\tilde c_1,\dots,\tilde c_n)}d\tilde c_i
\end{align*}
%equivalently,
%\begin{align}\label{eqn:omegarel}
%\sum_{i=1}^n \ab(\partial_{\square_i}\llparenthesis x:T \mid_\emptyset l\rrparenthesis_{(\tilde c_1,\dots,\tilde c_n)} - \partial_{\square_i}\llparenthesis x:T \mid_\emptyset r\rrparenthesis_{(\tilde c_1,\dots,\tilde c_n)} ) d\tilde c_i = 0,
%\end{align}
for $x:T \vdash l \approx r$ in $E$.

\begin{lemma}
$\epsilon$ is an epimorphism.
\end{lemma}
\begin{IEEEproof}
Let $f  = (x:T \mid_E t)\in \Mor(\bfC)$.
For $s = \sum_{i=1}^n \partial_{\square_i}\llparenthesis x:T \mid_\emptyset t \rrparenthesis_{\ab(\tilde c_1,\dots,\tilde c_n)}\gen{c_i}$, we have
\begin{align*}
\epsilon(s) &= \sum_{i=1}^n \partial_{\square_i}\llparenthesis x : T \mid_\emptyset t \rrparenthesis_{(\tilde c_1,\dots,\tilde c_n)}d\tilde c_i\\
 &= d (\llparenthesis x:T\mid_\emptyset t \rrparenthesis \cdot (\tilde c_1,\dots,\tilde c_n)) = df.
\end{align*}
\end{IEEEproof}

\begin{lemma}
$\ker \epsilon = \im \delta_1$.
\end{lemma}
\begin{IEEEproof}
If $\epsilon(s) = 0$, then $s$ can be written as a linear combination of elements of form
\[
\sum_{i=1}^n \ab(\partial_{\square_i}\llparenthesis x:T \mid_\emptyset l\rrparenthesis_{(\tilde c_1,\dots,\tilde c_n)} \gen{c_i} - \partial_{\square_i}\llparenthesis x:T \mid_\emptyset r\rrparenthesis_{(\tilde c_1,\dots,\tilde c_n)}  \gen{c_i}) 
\]
for $x:T \vdash l \approx r$ in $E$ and the above equals
$\delta_1(\gen{x:T\vdash l \approx r})$.
\end{IEEEproof}

%Let $B$ be a homotopy basis of $P^{(2)}(G_1(\Sigma,R))$.
We construct a $\cU_\bfC$-linear map
\(
\delta_2 : \cU_\bfC B \to \cU_\bfC E
\)
such that $\ker \delta_1 = \im \delta_2$.
For a path $p = (\omega_1,\vareqn_1); \dots; (\omega_m,\vareqn_m)$, we define $\varphi_1(p) \in \cU_\bfC E$ as
\[
\varphi_1(p) = \sum_{i=1}^m \varepsilon_i\cdot \partial(\omega_i)\gen{e_i'}
\]
where $e_i' = e_i, \varepsilon_i = 1$ if $e_i \in E$ and $e_i' = e_i^{-1}, \varepsilon_i = -1$ if $e_i \in E^{-1}$.
Then, we define $\delta_2\ab(\gen{p\parallel q})$ as
\[
\delta_2\ab(\gen{p\parallel q}) = \varphi_1\circ (\psi_{2,\mathrm L} - \psi_{2,\mathrm R})\ab(\gen{p \parallel q}) = \varphi_1(p) - \varphi_1(q).
\]

\begin{lemma}\label{lem:varphi0}
For any $\tau,\tau'\in\Mor(\Cl(\Sigma))$ and $\omega$ satisfying $\tau = \omega\cdot_\Sigma \tau'$, we have
\begin{align*}
\varphi_0(\tau) &= \sum_{i=1}^n \partial_{\square_i}(\omega)_{(\tilde c_1,\dots,\tilde c_n,[\tau']_E)}\gen{c_i}
+ \partial(\omega)_{([\tau']_E)}\varphi_0(\tau').
\end{align*}
\end{lemma}
\begin{IEEEproof}
By
$\llparenthesis \tau \rrparenthesis = [ \iota_1,\dots,\iota_n, \llparenthesis \tau' \rrparenthesis ]\circ \omega$.
%we have
%\begin{align*}
% & \partial_{\square_i}\llparenthesis \tau\rrparenthesis_{(\tilde c_1,\dots,\tilde c_n)} \\
%&=
%\sum_{i=1}^n\partial_{\square_i}(\omega)_{(\tilde c_1,\dots,\tilde c_n,\tau)} + \partial(\omega)_{(\tau)}\circ\partial_{\square_i}\llparenthesis \tau \rrparenthesis_{(\tilde c_1,\dots,\tilde c_n)}
%\end{align*}
%and then the desired equality holds by the definition of $\varphi_0$.
\end{IEEEproof}

\begin{lemma}
For a path $p$ from $\tau$ to $\tau'$ in $G_1(\Sigma,E)$,
$\delta_1\varphi_1(p) = \varphi_0(\tau) - \varphi_0(\tau')$.
\end{lemma}
\begin{IEEEproof}
We prove by induction on the length $k$ of $p$.
If $k=0$, then we have $u=v$ and obviously $\delta_1\varphi_1(p) = 0$.

If $k>0$, suppose $p = (\omega,\vareqn);p'$ for $\vareqn = (x:T \vdash l \approx r)$.
Then, we have
\begin{align*}
\delta_1\varphi_1(p)
 &= \partial(\omega)(\varphi_0(x:T \mid_\emptyset l) - \varphi_0(x:T \mid_\emptyset r))\\
&\quad + \varphi_0(\omega\cdot_\Sigma (x:T \mid_\emptyset r)) - \varphi_0(\tau')\\
&= \varphi_0(\omega\cdot_\Sigma (x:T\mid_\emptyset l)) - \varphi_0(\omega\cdot_\Sigma (x:T\mid_\emptyset r))\\
&\quad + \varphi_0(\omega\cdot_\Sigma (x:T\mid_\emptyset r)) - \varphi_0(\tau')\\
&= \varphi_0(\tau) - \varphi_0(\tau').
\end{align*}
Here, the second equality is implied from Lemma \ref{lem:varphi0}.
\end{IEEEproof}

For $\varpi \in P(G_2)$, define $\varphi_2(\varpi) \in \cU_\bfC B$ by 
$\varphi_2(1_p) = 0$ and
\begin{align*}
&\varphi_2((\omega,r_1,p\parallel q,r_2);\varpi')\\
&=\begin{cases}
\varphi_2(\varpi') & \text{if $p\parallel q \notin B\cup B^{-1}$},\\
\varphi_2(\varpi') + \partial(\omega)\gen{p\parallel q} & \text{if $p\parallel q \in B$},\\
\varphi_2(\varpi') - \partial(\omega)\gen{p\parallel q} & \text{if $p\parallel q \in B^{-1}$}.
\end{cases}
\end{align*}
The following follow from a simple computation.
\begin{lemma}
$\delta_1\delta_2 = 0$.
\end{lemma}
\begin{lemma}
$\delta_2\varphi_2(\varpi) = \varphi_1(p) - \varphi_1(q)$ if $\varpi\in P(G_2)$ is a path from $p$ to $q$.
\end{lemma}

We define a ringoid $\cQ_{\mathrm{rd}}$ as $\Ob(\cQ_{\mathrm{rd}}) = \Ob(\cU_\bfC)$ and $\Hom_{\cQ_{\mathrm{rd}}}(f,f) = \bbQ$ ($f \in \Mor(\bfC)$), $\Hom_{\cQ_{\mathrm{rd}}}(f,g) = \emptyset$ ($f \neq g \in \Mor(\bfC)$), and the composition is the multiplication in $\bbQ$.
It is obvious that any $\cU_\bfC$-modules are also $\cQ_{\mathrm{rd}}$-modules.

\begin{theorem}
$\ker \delta_1 = \im \delta_2$.
\end{theorem}
\begin{IEEEproof}
It suffices to construct two $\cQ_{\mathrm{rd}}$-linear maps
 $\eta_1 : \cU_\bfC \Sigma \to \cU_\bfC E$, $\eta_2 : \cU_\bfC E \to \cU_\bfC B$ such that
\[
\delta_2\eta_2(s) + \eta_1\delta_1(s) = ns
\]
for any $s \in \cU_\bfC E$ because then we have $\delta_2(\eta_2((1/n)s)) = s$ for any $s \in \ker\delta_1$.

For each $f : T_1 \to T_2$ in $\Mor(\bfC)$, we choose $\tau : T_1 \to T_2$ in $\Cl(\Sigma)$ such that $[\tau]_E = f$.
We call such $\tau$ the \emph{representative} of $f$.
Also, for each $\tau$ in $\Cl(\Sigma)$, we choose a path $p(\tau)$ in $G_1(\Sigma,R)$ from $\tau$ to the representative of $[\tau]_E$.

For $\omega :\Cl(\{T \to T'\}) \to \Cl(\Sigma \uplus \{\square : T_i \to T_i'\})$ and $c_i \in \Sigma$, we define $\eta_1(\partial(\omega)\gen{c_i}) $ as follows.
Suppose $\omega = \llparenthesis x:T \mid_\emptyset t : T \rrparenthesis$.
Let $t' = t[y/\square]$ for a fresh variable $y$.
If the representative of $\ab(z:  T \times (T_i \to T_i') \mid_E t'[\pr_1z/x,\pr_2z/y])$ is written as
\[
\ab(z:T\times (T_{i} \to T_i') \mid_\emptyset t''[\pr_1z/x,\pr_2z/y])
\]
for some term $t''$ with free variables $x,y$, then
\[
\eta_1\ab(\partial(\omega)\gen{c_i}) = \varphi_1\ab(p\ab(x:T \mid_\emptyset t''[c_i/y])).
\]

We define paths $p_i(\omega,\vareqn)$ and $q_i(\omega,\vareqn)$ ($i=1,\dots,n$) for $\vareqn = (x:T \vdash l \approx r)$ as follows.
Suppose $\omega\cdot_\Sigma (x:T \mid_\emptyset l) = (x:T \mid_\emptyset t)$ and $\omega\cdot_\Sigma (x:T \mid_\emptyset r) = (x:T \mid_\emptyset t')$.
Let $l'_i = t[y_i/c_i]$, $r'_i = t'[y_i/c_i]$ for fresh variable $y_i$.
Also, let $\ab(z:T \times (T_i \to T_i')\mid_\emptyset l''_i[\pr_1z/x,\pr_2/y_i])$ be the representative of $\ab(z:T\times (T_i \to T_i')\mid_E l'_i[\pr_1z/x,\pr_2z/y_i])$ and $\ab(z:T\times(T_i\to T_i')\mid_\emptyset r''_i[\pr_1z/x,\pr_2z/y_i])$ be the representative of $\ab(z:T\times(T_i\to T_i') \mid_E r'_i[\pr_1z/x,\pr_2z/y_i])$.

We define $p_i(\omega,\vareqn)$ and $q_i(\omega,\vareqn)$ as
\begin{align*}
p_i(\omega,\vareqn) = (\omega,\vareqn); p'_i(\omega,e),\quad
q_i(\omega,\vareqn) = p\ab(x:T\mid_\emptyset l''_i[c_i/y_i]).
\end{align*}
where $p'_i(\omega,\vareqn) = p\ab(x:T\mid_\emptyset r''_i[c_i/y_i])$.
Also, for any $p\parallel q \in P^{(2)}(G_1(\Sigma,E))$, we choose a path $\varpi(p,q)$ in $G_2(B)$ from $p$ to $q$.

Then, define $\eta_2 : \cU_\bfC E \to \cU_\bfC B$ as
\[
\eta_2(\partial(\omega)\gen{\vareqn}) =\sum_{i=1}^n\varphi_2(\varpi(p_i(\omega,\vareqn), q_i(\omega,\vareqn))).
\]
By the definition of $\eta_1$ and $\eta_2$,
\begin{align*}
\delta_2\eta_2(\partial(\omega)\gen{\vareqn})
&= \sum_{i=1}^n \delta_2\varphi_2(\varpi(p_i(\omega,\vareqn), q_i(\omega,\vareqn)))\\
&= \sum_{i=1}^n \varphi_1(p_i(\omega,\vareqn)) - \varphi_1(q_i(\omega,\vareqn))\\
&= n\partial(\omega)\gen{\vareqn} +\sum_{i=1}^n \varphi_1(p_i'(\omega,e)) - \varphi_1(q_i(\omega,e)),
\end{align*}
\begin{align*}
&\eta_1\delta_1(\partial(\omega)\gen{\vareqn})\\
&= \eta_1(\partial(\omega)\varphi_0(x:T\mid_\emptyset l) - \partial(\omega)\varphi_0(x:T\mid_\emptyset r))\\
&= \eta_1\biggl(\sum_{i=1}^n  \partial_i \llparenthesis x:T\mid_\emptyset l \rrparenthesis_{(\tilde c_1,\dots,\tilde c_n)}\gen{c_i}\\
&\quad\qquad - \sum_{i=1}^n\partial_i \llparenthesis x:T\mid_\emptyset r \rrparenthesis_{(\tilde c_1,\dots,\tilde c_n)}\gen{c_i}\biggr)\\
&=\sum_{i=1}^n \varphi_1(q_i(\omega,e)) - \varphi_1(p_i'(\omega,e)).
\end{align*}
Thus, we conclude 
\(
\delta_2\eta_2(\partial(\omega)\gen{\vareqn}) + \eta_1\delta_1(\partial(\omega)\gen{\vareqn}) = n\partial(\omega)\gen{\vareqn}.
\)
\end{IEEEproof}

Now, we have proved that the sequence (\ref{eqn:resolution_omega}) is exact.
Consider the chain complex
\[
\cQ \otimes_{\cU_\bfC} \cU_\bfC B \xrightarrow{\cQ \otimes \delta_2} \cQ \otimes_{\cU_\bfC} \cU_\bfC E \xrightarrow{\cQ \otimes \delta_1}\cQ \otimes_{\cU_\bfC} \cU_\bfC \Sigma 
\]

Let $H_0(\bfC) = \bbQ \Sigma/ \im (\cQ \otimes \delta_1)$, $H_1(\bfC) = \ker(\cQ \otimes \delta_1)/\im(\cQ \otimes \delta_2)$.

\begin{theorem} Let $e(E) = \dim(H_1(\bfC))-\dim(H_0(\bfC))+\#\Sigma$ for $\bfC = \Cl(\Sigma,E)$. Then
$e(E) = \#E - \dim(\im(\cQ \otimes \delta_2)).$
\end{theorem}
\begin{IEEEproof}
Immediate from
\begin{align*}
&\dim(H_1(\bfC)) - \dim(H_0(\bfC))\\
&= \dim(\ker(\cQ \otimes \delta_1)) - \dim(\im(\cQ \otimes \delta_2)) - \#\Sigma\\
&\quad+ \dim(\im(\cQ \otimes \delta_1))
\end{align*}
and $\dim(\ker (\cQ \otimes \delta_1))+\dim(\im(\cQ \otimes \delta_1)) = \#E$ (rank-nullity theorem).
\end{IEEEproof}

Note that if $E$ and $E'$ are equivalent, then $\Cl(\Sigma,E) \cong \Cl(\Sigma,E')$, so $e(E) = e(E')$.

The integer $e(E)$ is computable if $E$ has FDT and the homotopy basis $B$ is explicitly given.
In particular, if $E=R$ is a finite complete PRS and we take $B = C(\Sigma,E)$, then
we can check that the second boundary matrix $D_2(R)$ is a matrix representation of $\cQ \otimes \delta_2$.
Since $\dim(\im(\cQ \otimes \delta_2)) = \rank(D_2(R))$, we obtain Theorem \ref{thm:main}, $\#R - \rank(D_2(R)) = e(R) = e(E') \le \#E'$ for any $E'$ equivalent to $R$.

\section{Concluding remarks and future work}

We proved that, given an equation system, a lower bound on the number of higher-order equations is obtained by a homological approach to the equation system.
If the equation system has a complete PRS, then the lower bound can be computed by simple matrix operations.

Even without a complete PRS, there is still a chance to compute a lower bound.
If one finds a finite homotopy basis for the equation system, then a similar computation gives a lower bound.
One of the interesting future works is to investigate how to find a homotopy basis of an equation system like one including commutativity.

In \cite{ikebuchi21}, Ikebuchi showed that there is a homological necessary condition of the $E$-unifiability for a first-order equation system $E$.
The author believes that this result also can be extended to higher-order equations.

\section*{Acknowledgment}
The author would like to thank Haynes Miller for fruitful discussions on the background theory of this work.
This work was supported by JSPS KAKENHI Grant Number 24K20758.

\bibliographystyle{IEEEtran}
\bibliography{lics25}

\end{document}

%% file: common.tex
\theoremstyle{definition}
\newtheorem{theorem}{Theorem}
\newtheorem{corollary}[theorem]{Corollary}
\newtheorem{lemma}[theorem]{Lemma}

\newtheorem{proposition}[theorem]{Proposition}
\newtheorem{definition}[theorem]{Definition}

\newtheorem{example}[theorem]{Example}

\def\cE{{\mathcal E}}

\def\cP{{\mathcal P}}
\def\cQ{{\mathcal Q}}
\def\cR{{\mathcal R}}
\def\cS{{\mathcal S}}

\def\cU{{\mathcal U}}

\def\bbQ{{\mathbb Q}}

\def\bfC{{\mathbf C}}
\def\bfD{{\mathbf D}}

\def\bfT{{\mathbf T}}

\DeclareMathOperator{\im}{im}
\newcommand{\Hom}{\mathrm{Hom}}

\DeclareMathOperator{\Tor}{Tor}

\DeclareMathOperator{\Alg}{Alg}

\newcommand{\Mor}{\mathrm{Mor}}

\newcommand{\Id}{\mathrm{id}}
\newcommand{\Ob}{\mathrm{Ob}}
\newcommand{\Op}{\mathrm{op}}

\newcommand{\catab}{\mathbf{Ab}}
\newcommand{\catset}{\mathbf{Set}}

\newcommand{\catfam}{\mathbf{Fam}}
\newcommand{\catcfam}{\mathbf{CFam}}

\newcommand{\catccc}{\mathbf{CCC}}

\newcommand{\catmodules}[1]{#1\text{-}\mathbf{mod}}

\DeclareMathOperator{\rank}{rank}

\makeatletter
\newcommand*{\relrelbarsep}{.386ex}
\newcommand*{\relrelbar}{%
  \mathrel{%
    \mathpalette\@relrelbar\relrelbarsep
  }%
}
\newcommand*{\@relrelbar}[2]{%
  \raise#2\hbox to 0pt{$\m@th#1\relbar$\hss}%
  \lower#2\hbox{$\m@th#1\relbar$}%
}
\providecommand*{\rightrightarrowsfill@}{%
  \arrowfill@\relrelbar\relrelbar\rightrightarrows
}
\providecommand*{\leftleftarrowsfill@}{%
  \arrowfill@\leftleftarrows\relrelbar\relrelbar
}
\providecommand*{\xrightrightarrows}[2][]{%
  \ext@arrow 0359\rightrightarrowsfill@{#1}{#2}%
}
\providecommand*{\xleftleftarrows}[2][]{%
  \ext@arrow 3095\leftleftarrowsfill@{#1}{#2}%
}
\makeatother